\input harvmac
\let\includefigures=\iftrue
\let\useblackboard==\iftrue
\newfam\black

\includefigures
\message{If you do not have epsf.tex (to include figures),}
\message{change the option at the top of the tex file.}
\input epsf
\def\figin{\epsfcheck\figin}\def\figins{\epsfcheck\figins}
\def\epsfcheck{\ifx\epsfbox\UnDeFiNeD
\message{(NO epsf.tex, FIGURES WILL BE IGNORED)}
\gdef\figin##1{\vskip2in}\gdef\figins##1{\hskip.5in}
\else\message{(FIGURES WILL BE INCLUDED)}%
\gdef\figin##1{##1}\gdef\figins##1{##1}\fi}
\def\DefWarn#1{}
\def\figinsert{\goodbreak\midinsert}
\def\ifig#1#2#3{\DefWarn#1\xdef#1{fig.~\the\figno}
\writedef{#1\leftbracket fig.\noexpand~\the\figno}%
\figinsert\figin{\centerline{#3}}\medskip\centerline{\vbox{
\baselineskip12pt\advance\hsize by -1truein
\noindent\footnotefont{\bf Fig.~\the\figno:} #2}}
\endinsert\global\advance\figno by1}
\else
\def\ifig#1#2#3{\xdef#1{fig.~\the\figno}
\writedef{#1\leftbracket fig.\noexpand~\the\figno}%
\global\advance\figno by1} \fi
\def\id{{1 \kern-.28em {\rm l}}}

\def\K3{{\bf K3}}
\def\journal#1&#2(#3){\unskip, \sl #1\ \bf #2 \rm(19#3) }
\def\andjournal#1&#2(#3){\sl #1~\bf #2 \rm (19#3) }

\def\bar{\overline}
\def\hat{\widehat}
\def\ie{{\it i.e.}}
\def\eg{{\it e.g.}}

\def\frac#1#2{{#1\over#2}}

\def\half{\frac12}

\def\inbar{\,\vrule height1.5ex width.4pt depth0pt}
\def\IC{\relax\hbox{$\inbar\kern-.3em{\rm C}$}}
\def\IR{\relax{\rm I\kern-.18em R}}
\def\IP{\relax{\rm I\kern-.18em P}}

%
%

%
\catcode`\@=11
\def\slash#1{\mathord{\mathpalette\c@ncel{#1}}}
\overfullrule=0pt

\def\FF{{\cal F}}

\def\LL{{\cal L}}
\def\MM{{\cal M}}
\def\NN{{\cal N}}

\def\SS{{\cal S}}

\def\underrel#1\over#2{\mathrel{\mathop{\kern\z@#1}\limits_{#2}}}

\catcode`\@=12


%

\def\exp{{\rm exp}}


\def\ie{{\it i.e.}}
\def\eg{{\it e.g.}}



\lref\ElShowkCM{
  S.~El-Showk and M.~Guica,
  ``Kerr/CFT, dipole theories and nonrelativistic CFTs,''
JHEP {\bf 1212}, 009 (2012).
[arXiv:1108.6091 [hep-th]].
}

\lref\CompereJK{
  G.~Compère,
  ``The Kerr/CFT correspondence and its extensions,''
Living Rev.\ Rel.\  {\bf 15}, 11 (2012), [Living Rev.\ Rel.\  {\bf 20}, no. 1, 1 (2017)].
[arXiv:1203.3561 [hep-th]].
}

\lref\deBoerGYT{
  J.~de Boer, H.~Ooguri, H.~Robins and J.~Tannenhauser,
  ``String theory on AdS(3),''
JHEP {\bf 9812}, 026 (1998).
[hep-th/9812046].
}

\lref\GiveonKU{
  A.~Giveon and A.~Pakman,
  ``More on superstrings in AdS(3) x N,''
JHEP {\bf 0303}, 056 (2003).
[hep-th/0302217].
}

\lref\BzowskiPCY{
  A.~Bzowski and M.~Guica,
  ``The holographic interpretation of $J \bar T$-deformed CFTs,''
[arXiv:1803.09753 [hep-th]].
}

\lref\GiveonFU{
  A.~Giveon, M.~Porrati and E.~Rabinovici,
  ``Target space duality in string theory,''
Phys.\ Rept.\  {\bf 244}, 77 (1994).
[hep-th/9401139].
}

\lref\ParsonsSI{
  J.~Parsons and S.~F.~Ross,
  ``Strings in extremal BTZ black holes,''
JHEP {\bf 0904}, 134 (2009).
[arXiv:0901.3044 [hep-th]].
}

\lref\BarsSR{
  I.~Bars and K.~Sfetsos,
  ``Conformally exact metric and dilaton in string theory on curved space-time,''
Phys.\ Rev.\ D {\bf 46}, 4510 (1992).
[hep-th/9206006].
}

\lref\DetournayFZ{
  S.~Detournay, D.~Orlando, P.~M.~Petropoulos and P.~Spindel,
  ``Three-dimensional black holes from deformed anti-de Sitter,''
JHEP {\bf 0507}, 072 (2005).
[hep-th/0504231].
}

\lref\AzeyanagiZD{
  T.~Azeyanagi, D.~M.~Hofman, W.~Song and A.~Strominger,
  ``The Spectrum of Strings on Warped $AdS_3 \times S^3$,''
JHEP {\bf 1304}, 078 (2013).
[arXiv:1207.5050 [hep-th]].
}

\lref\SmirnovLQW{
  F.~A.~Smirnov and A.~B.~Zamolodchikov,
  ``On space of integrable quantum field theories,''
Nucl.\ Phys.\ B {\bf 915}, 363 (2017).
[arXiv:1608.05499 [hep-th]].
}

\lref\CavagliaODA{
  A.~Cavaglià, S.~Negro, I.~M.~Szécsényi and R.~Tateo,
  ``$T \bar{T}$-deformed 2D Quantum Field Theories,''
JHEP {\bf 1610}, 112 (2016).
[arXiv:1608.05534 [hep-th]].
}

\lref\McGoughLOL{
  L.~McGough, M.~Mezei and H.~Verlinde,
  ``Moving the CFT into the bulk with $ T\overline{T} $,''
JHEP {\bf 1804}, 010 (2018).
[arXiv:1611.03470 [hep-th]].
}

\lref\GiveonNIE{
  A.~Giveon, N.~Itzhaki and D.~Kutasov,
  ``$ T\bar{T} $ and LST,''
JHEP {\bf 1707}, 122 (2017).
[arXiv:1701.05576 [hep-th]].
}

\lref\PolchinskiRQ{
  J.~Polchinski,
  ``String theory. Vol. 1: An introduction to the bosonic string,''
}

\lref\GiveonMYJ{
  A.~Giveon, N.~Itzhaki and D.~Kutasov,
  ``A solvable irrelevant deformation of AdS$_{3}$/CFT$_{2}$,''
JHEP {\bf 1712}, 155 (2017).
[arXiv:1707.05800 [hep-th]].
}

\lref\MaldacenaHW{
  J.~M.~Maldacena and H.~Ooguri,
  ``Strings in AdS(3) and SL(2,R) WZW model 1: The Spectrum,''
J.\ Math.\ Phys.\  {\bf 42}, 2929 (2001).
[hep-th/0001053].
}

\lref\ShyamZNQ{
  V.~Shyam,
  ``Background independent holographic dual to $T\bar{T}$ deformed CFT with large central charge in 2 dimensions,''
JHEP {\bf 1710}, 108 (2017).
[arXiv:1707.08118 [hep-th]].
}

\lref\AsratTZD{
  M.~Asrat, A.~Giveon, N.~Itzhaki and D.~Kutasov,
  ``Holography Beyond AdS,''
[arXiv:1711.02690 [hep-th]].
}

\lref\GiribetIMM{
  G.~Giribet,
  ``$T\bar{T}$-deformations, AdS/CFT and correlation functions,''
JHEP {\bf 1802}, 114 (2018).
[arXiv:1711.02716 [hep-th]].
}

\lref\KrausXRN{
  P.~Kraus, J.~Liu and D.~Marolf,
  ``Cutoff AdS$_3$ versus the $T\bar{T}$ deformation,''
[arXiv:1801.02714 [hep-th]].
}

\lref\CardySDV{
  J.~Cardy,
  ``The $T\overline T$ deformation of quantum field theory as a stochastic process,''
[arXiv:1801.06895 [hep-th]].
}

\lref\CottrellSKZ{
  W.~Cottrell and A.~Hashimoto,
  ``Comments on $T \bar T$ double trace deformations and boundary conditions,''
[arXiv:1801.09708 [hep-th]].
}

\lref\AharonyVUX{
  O.~Aharony and T.~Vaknin,
  ``The TT* deformation at large central charge,''
[arXiv:1803.00100 [hep-th]].
}

\lref\DubovskyDLK{
  S.~Dubovsky,
  ``A Simple Worldsheet Black Hole,''
[arXiv:1803.00577 [hep-th]].
}

\lref\BonelliKIK{
  G.~Bonelli, N.~Doroud and M.~Zhu,
  ``$T\bar T$-deformations in closed form,''
[arXiv:1804.10967 [hep-th]].
}

\lref\BaggioGCT{
  M.~Baggio and A.~Sfondrini,
  ``Strings on NS-NS Backgrounds as Integrable Deformations,''
[arXiv:1804.01998 [hep-th]].
}

\lref\ChakrabortyKPR{
  S.~Chakraborty, A.~Giveon, N.~Itzhaki and D.~Kutasov,
  ``Entanglement Beyond $\rm AdS$,''
[arXiv:1805.06286 [hep-th]].
}

\lref\GuicaLIA{
  M.~Guica,
  ``An integrable Lorentz-breaking deformation of two-dimensional CFTs,''
[arXiv:1710.08415 [hep-th]].
}

\lref\BzowskiPCY{
  A.~Bzowski and M.~Guica,
  ``The holographic interpretation of $J \bar T$-deformed CFTs,''
[arXiv:1803.09753 [hep-th]].
}

\lref\IRownNW{
  J.~D.~Brown and M.~Henneaux,
  ``Central Charges in the Canonical Realization of Asymptotic Symmetries: An Example from Three-Dimensional Gravity,''
Commun.\ Math.\ Phys.\  {\bf 104}, 207 (1986)..
}

\lref\GiveonCGS{
  A.~Giveon and D.~Kutasov,
  ``Supersymmetric Renyi entropy in CFT$_{2}$ and AdS$_{3}$,''
JHEP {\bf 1601}, 042 (2016).
[arXiv:1510.08872 [hep-th]].
}

\lref\KutasovXQ{
  D.~Kutasov and A.~Schwimmer,
  ``Universality in two-dimensional gauge theory,''
Nucl.\ Phys.\ B {\bf 442}, 447 (1995).
[hep-th/9501024].
}

\lref\GiveonNS{
  A.~Giveon, D.~Kutasov and N.~Seiberg,
  ``Comments on string theory on AdS(3),''
Adv.\ Theor.\ Math.\ Phys.\  {\bf 2}, 733 (1998).
[hep-th/9806194].
}

\lref\KutasovXU{
  D.~Kutasov and N.~Seiberg,
  ``More comments on string theory on AdS(3),''
JHEP {\bf 9904}, 008 (1999).
[hep-th/9903219].
}

\lref\GiveonUP{
  A.~Giveon and D.~Kutasov,
  ``Notes on AdS(3),''
Nucl.\ Phys.\ B {\bf 621}, 303 (2002).
[hep-th/0106004].
}

\lref\ArgurioTB{
  R.~Argurio, A.~Giveon and A.~Shomer,
  ``Superstrings on AdS(3) and symmetric products,''
JHEP {\bf 0012}, 003 (2000).
[hep-th/0009242].
}

\lref\GiveonMI{
  A.~Giveon, D.~Kutasov, E.~Rabinovici and A.~Sever,
  ``Phases of quantum gravity in AdS(3) and linear dilaton backgrounds,''
Nucl.\ Phys.\ B {\bf 719}, 3 (2005).
[hep-th/0503121].
}

\lref\MotlTH{
  L.~Motl,
  ``Proposals on nonperturbative superstring interactions,''
[hep-th/9701025].
}

\lref\DijkgraafVV{
  R.~Dijkgraaf, E.~P.~Verlinde and H.~L.~Verlinde,
  ``Matrix string theory,''
Nucl.\ Phys.\ B {\bf 500}, 43 (1997).
[hep-th/9703030].
}

\lref\GiveonCGS{
  A.~Giveon and D.~Kutasov,
  ``Supersymmetric Renyi entropy in CFT$_{2}$ and AdS$_{3}$,''
JHEP {\bf 1601}, 042 (2016).
[arXiv:1510.08872 [hep-th]].
}

\lref\KutasovXB{
  D.~Kutasov,
  ``Geometry on the Space of Conformal Field Theories and Contact Terms,''
Phys.\ Lett.\ B {\bf 220}, 153 (1989).
}

\lref\GiveonZM{
  A.~Giveon, D.~Kutasov and O.~Pelc,
  ``Holography for noncritical superstrings,''
JHEP {\bf 9910}, 035 (1999).
[hep-th/9907178].
}

\lref\ZamolodchikovBD{
  A.~B.~Zamolodchikov and V.~A.~Fateev,
  ``Operator Algebra and Correlation Functions in the Two-Dimensional Wess-Zumino SU(2) x SU(2) Chiral Model,''
Sov.\ J.\ Nucl.\ Phys.\  {\bf 43}, 657 (1986), [Yad.\ Fiz.\  {\bf 43}, 1031 (1986)].
}

\lref\TeschnerFT{
  J.~Teschner,
  ``On structure constants and fusion rules in the SL(2,C) / SU(2) WZNW model,''
Nucl.\ Phys.\ B {\bf 546}, 390 (1999). [hep-th/9712256]; A. B. Zamolodchikov and Al. B.
Zamolodchikov, unpublished.
}

\lref\KlemmDF{
  A.~Klemm and M.~G.~Schmidt,
  ``Orbifolds by Cyclic Permutations of Tensor Product Conformal Field Theories,''
Phys.\ Lett.\ B {\bf 245}, 53 (1990).
}

\lref\FuchsVU{
  J.~Fuchs, A.~Klemm and M.~G.~Schmidt,
  ``Orbifolds by cyclic permutations in Gepner type superstrings and in the corresponding Calabi-Yau manifolds,''
Annals Phys.\  {\bf 214}, 221 (1992).
}

\lref\WakimotoGF{
  M.~Wakimoto,
  ``Fock representations of the affine lie algebra A1(1),''
Commun.\ Math.\ Phys.\  {\bf 104}, 605 (1986).
}

\lref\BernardIY{
  D.~Bernard and G.~Felder,
  ``Fock Representations and BRST Cohomology in SL(2) Current Algebra,''
Commun.\ Math.\ Phys.\  {\bf 127}, 145 (1990).
}

\lref\BershadskyIN{
  M.~Bershadsky and D.~Kutasov,
  ``Comment on gauged WZW theory,''
Phys.\ Lett.\ B {\bf 266}, 345 (1991).
}

\lref\ElitzurRT{
  S.~Elitzur, A.~Giveon, D.~Kutasov and E.~Rabinovici,
  ``From big bang to big crunch and beyond,''
JHEP {\bf 0206}, 017 (2002).
[hep-th/0204189].
}

\lref\CrapsII{
  B.~Craps, D.~Kutasov and G.~Rajesh,
  ``String propagation in the presence of cosmological singularities,''
JHEP {\bf 0206}, 053 (2002).
[hep-th/0205101].
}

\lref\ZamolodchikovCE{
  A.~B.~Zamolodchikov,
  ``Expectation value of composite field T anti-T in two-dimensional quantum field theory,''
[hep-th/0401146].
}

\lref\CaselleDRA{
  M.~Caselle, D.~Fioravanti, F.~Gliozzi and R.~Tateo,
  ``Quantisation of the effective string with TBA,''
JHEP {\bf 1307}, 071 (2013).
[arXiv:1305.1278 [hep-th]].
}

\lref\DattaTHY{
  S.~Datta and Y.~Jiang,
  ``$T\bar{T}$ deformed partition functions,''
[arXiv:1806.07426 [hep-th]].
}

\lref\IsraelVV{
  D.~Israel, C.~Kounnas, D.~Orlando and P.~M.~Petropoulos,
  ``Electric/magnetic deformations of S**3 and AdS(3), and geometric cosets,''
Fortsch.\ Phys.\  {\bf 53}, 73 (2005).
[hep-th/0405213].
}

\lref\DubovskyIRA{
  S.~Dubovsky, V.~Gorbenko and M.~Mirbabayi,
  ``Natural Tuning: Towards A Proof of Concept,''
JHEP {\bf 1309}, 045 (2013).
[arXiv:1305.6939 [hep-th]].
}

\lref\DubovskyCNJ{
  S.~Dubovsky, V.~Gorbenko and M.~Mirbabayi,
  ``Asymptotic fragility, near AdS$_{2}$ holography and $ T\overline{T} $,''
JHEP {\bf 1709}, 136 (2017).
[arXiv:1706.06604 [hep-th]].
}

\lref\ApoloQPQ{
  L.~Apolo and W.~Song,
  ``Strings on warped AdS$_3$ via $T\bar{J}$ deformations,''
[arXiv:1806.10127 [hep-th]].
}

\lref\DetournayRH{
  S.~Detournay, D.~Israel, J.~M.~Lapan and M.~Romo,
  ``String Theory on Warped $AdS_{3}$ and Virasoro Resonances,''
JHEP {\bf 1101}, 030 (2011).
[arXiv:1007.2781 [hep-th]].
}

\Title{} {\centerline{$J\bar{T}$ deformed $CFT_2$ and String Theory}}

\bigskip
\centerline{\it Soumangsu Chakraborty${}^{1}$, Amit Giveon${}^{1}$ and David Kutasov${}^{2}$}
\bigskip
\smallskip
\centerline{${}^{1}$Racah Institute of Physics, The Hebrew
University} \centerline{Jerusalem 91904, Israel}
\smallskip
\centerline{${}^2$EFI and Department of Physics, University of
Chicago} \centerline{5640 S. Ellis Av., Chicago, IL 60637, USA }

\smallskip

\vglue .3cm

\bigskip

\bigskip
\noindent
We study two dimensional conformal field theory with a left-moving conserved current $J$, perturbed by an irrelevant, Lorentz symmetry breaking operator with the quantum numbers of $J\bar T$, using a combination of field and string theoretic techniques. We show that the spectrum of the theory has some interesting features, which may shed light on systems of interest for holography and black hole physics.

\bigskip

\Date{6/18}


\newsec{Introduction}
Consider a two dimensional conformal field theory ($CFT_2$), which contains a conserved left-moving $U(1)$ current\foot{We will label the (Euclidean) space on which the theory lives by complex coordinates $x$, $\bar x$. The conservation equation for the current $J$ is thus $\partial_{\bar x} J=0$.} $J(x)$. Now, deform the theory by adding to its Lagrangian the term
\eqn\aaa{\LL_{\rm int}=\mu J(x)\bar T(\bar x),
}
where $\bar T$ is the anti-holomorphic component of the stress-tensor, $\bar T=T_{\bar x\bar x}$.
More precisely, \aaa\ is the form of the deformation for small $\mu$; we will describe the form at finite $\mu$ later. In particular, we will argue that one can define the theory in such a way that at every point in the space of theories labeled by $\mu$ \aaa, there is a holomorphic current $J(x)$ that satisfies $\partial_{\bar x} J(x)=0$, and changing $\mu$ by an infinitesimal amount $\delta\mu$ corresponds to adding to the Lagrangian an infinitesimal version of \aaa, $\delta\LL_{\rm int}=\delta\mu J(x)\bar T(\bar x)$, where the operators $J$, $\bar T$ are defined in the theory with coupling $\mu$.

Superficially, the theory \aaa\ is problematic. The coupling $\mu$ has left and right-moving scaling dimension $(0,-1)$; hence, the theory is not Lorentz invariant. Moreover, since $\mu$ has negative mass dimension, it goes to zero in the IR and grows in the UV. Thus, the description \aaa\ corresponds to a flow up the RG, which is usually ill-defined.

There are reasons to believe that in the case \aaa\ the situation is better. The authors of \refs{\SmirnovLQW,\CavagliaODA} analyzed the analogous model of $T\bar T$ deformed $CFT_2$, found that it is exactly solvable, and computed the spectrum on a cylinder (see also \refs{\ZamolodchikovCE\CaselleDRA-\DubovskyIRA} for earlier work). This model and closely related ones were further discussed in \refs{\McGoughLOL\GiveonNIE\DubovskyCNJ\GiveonMYJ\ShyamZNQ\GuicaLIA\AsratTZD\GiribetIMM\KrausXRN\CardySDV\CottrellSKZ\AharonyVUX\DubovskyDLK\BzowskiPCY\BonelliKIK\BaggioGCT\ChakrabortyKPR-\DattaTHY}.

In a sense, the model \aaa\ is an intermediate case between $T\bar T$ deformed $CFT_2$ and a model with left and right-moving currents $J(x)$, $\bar J(\bar x)$, with the marginal deformation $\LL_{\rm int}=\lambda J(x)\bar J(\bar x)$. The latter preserves both (left and right-moving) copies of Virasoro, and the left and right-moving $U(1)$ affine Lie algebras generated by $J$ and $\bar J$, respectively. The former breaks both copies of Virasoro to the $U(1)$'s corresponding to translations. As we will discuss, the model \aaa\ preserves the left-moving Virasoro, as well as the left-moving affine Lie algebra. The right-moving Virasoro symmetry is broken to a $U(1)$ corresponding to translations of $\bar x$.

In a recent paper \GuicaLIA, Guica studied $J\bar T$ deformed $CFT_2$ following the analysis of \refs{\SmirnovLQW,\CavagliaODA} of  the $T\bar T$ deformation. In particular, she analyzed the spectrum of the theory on a cylinder. In this note we will revisit this model and discuss its connection to holography. We will study the single trace $J\bar T$ deformation, generalizing the analysis of \refs{\GiveonNIE,\GiveonMYJ,\AsratTZD,\ChakrabortyKPR} of the $T\bar T$ deformation to this case. The double trace version was studied in \BzowskiPCY.

In the holographic context, our discussion is relevant to many of the basic examples of the $AdS_3/CFT_2$ correspondence, such as $AdS_3\times S^3\times T^4$, $AdS_3\times S^3\times S^3\times S^1$, and more general examples, such as those of \refs{\GiveonZM,\GiveonKU}.

Consider, for example, the background $AdS_3\times S^3\times T^4$ in type IIB string theory. It is obtained by studying the near-horizon geometry of $k$ $NS5$-branes wrapped around $S^1\times T^4$ and $N$ fundamental strings wrapped around $S^1$. The $T^4$ factor in the background corresponds to the space along the fivebranes and transverse to the strings, while the $S^3$ corresponds to the angular directions in the space transverse to both the strings and the fivebranes.

As discussed in \refs{\GiveonNS,\KutasovXU}, the boundary CFT that corresponds to string theory in this background contains $SU(2)\times U(1)^4$ left and right-moving affine Lie algebras. The first factor comes from the isometries of the $S^3$; the second is associated with momentum and winding on the $T^4$. We can take the $U(1)$ current $J(x)$ that figures in the construction \aaa\ to be either a $U(1)$ subgroup of $SU(2)_L$ or one of the left-moving $U(1)$'s associated with the $T^4$.

An important difference between the two has to do with supersymmetry. The original $AdS_3\times S^3\times T^4$ background preserves $(4,4)$ superconformal symmetry. If we take the $U(1)$ current $J(x)$ in \aaa\ to come from the $S^3$, the perturbation breaks SUSY to $(0,4)$, since in that case the current $J(x)$ is a bottom component of a superfield (it is an R-current). On the other hand, if we take the current to be one of the $U(1)$'s associated with the $T^4$, the deformation \aaa\ preserves $(4,4)$ SUSY (but, as mentioned above, not the right-moving conformal symmetry).

Another important issue in the context of holography is the existence of two versions of the deformation \aaa, associated with single and double trace deformations. It appeared already in the $T\bar T$ case \refs{\GiveonNIE,\GiveonMYJ}, and will play a role in our discussion here as well. Therefore, we next briefly review it.

One way to introduce this issue is the following. Consider a $CFT_2$ that has the symmetric product form $\MM^N/S_N$, where $\MM$ is a $CFT_2$ which we will refer to as the building block (or block, for short), of the symmetric product. In this theory there are two natural $T\bar T$ deformations. Denoting by $T_i$, $\bar T_i$, $i=1,\cdots, N$, the (anti) holomorphic stress-tensors of the $i$'th copy of $\MM$, the stress tensors of the symmetric product CFT are $T=\sum_i T_i$ and $\bar T=\sum_i\bar T_i$, and the two deformations are $T\bar T$ and $\sum_i T_i\bar T_i$. The first is the $T\bar T$ deformation of the full, symmetric product, theory; the second is the $T\bar T$ deformation of the block $\MM$.

If we now take the block $\MM$ to have a left-moving conserved current $J(x)$, we can repeat the discussion of the previous paragraph for the deformation \aaa. There are again two different deformations of this type -- the $J\bar T$ deformation of the full, symmetric product, CFT, and the $J\bar T$ deformation of the block $\MM$.

The boundary $CFT$ corresponding to string theory on $AdS_3$ is not quite a symmetric product $CFT$, however, it is closely related to one. For the purpose of our discussion, it is useful to note two things. The first is that, like in the symmetric product theory, string theory on $AdS_3$ has two different $T\bar T$ (and similarly $J\bar T$) deformations. For the $T\bar T$ case, this was discussed in \refs{\GiveonNIE,\GiveonMYJ}. For $J\bar T$ one can proceed similarly.

One deformation is obtained by using the vertex operators for $J(x)$ and $\bar T(\bar x)$ \KutasovXU, and multiplying them, as in \aaa. Since each of these operators is given by an integral over the worldsheet, the resulting deformation is a product of two such integrals, \ie\ it is a double trace deformation. The second deformation corresponds to adding to the spacetime Lagrangian the operator given by eq.  (6.5) in \KutasovXU. This is a supergravity mode and, from the boundary point of view, a single trace deformation.

These two operators appear to be very similar to the two different $J\bar T$ operators described above for the symmetric product. This resemblance is part of a much richer story about the relation between string theory on $AdS_3$ and symmetric products of the sort described above (see \eg\ \refs{\ArgurioTB,\GiveonMI}). This leads us to the second observation.

The background corresponding to the boundary $CFT$ in its Ramond sector (\ie\ with unbroken supersymmetry on the cylinder), is obtained by replacing $AdS_3$ by the $M=J=0$ BTZ black hole. The strings and fivebranes that create the background are mutually BPS in this case. Thus, their potential is flat. This means that there is a continuum of states corresponding to strings moving radially away from the fivebranes. These states are described by a symmetric product, as in matrix string theory \refs{\MotlTH,\DijkgraafVV}.

Therefore, in this sector of the Hilbert space, we expect the symmetric product picture to be correct (with the degree of the permutation group, $N$, being the number of strings creating the background), and the deformation given by eq. (6.5) in \KutasovXU\ to be just the $J\bar T$ deformation in the block $\MM$ corresponding to the CFT associated with one string. Thus, by studying the spectrum of string theory deformed by the operator (6.5) in \KutasovXU, we can get information about the spectrum of $J\bar T$ deformed $CFT_2$. One of our goals in this paper is to analyze the spectrum of string theory in the above background, and match it to that of $J\bar T$ deformed CFT.

The plan of this paper is as follows. In section {\it 2},
we review some aspects of string theory on $AdS_3$, including the worldsheet sigma model Lagrangian, its representation in terms of Wakimoto variables, the current algebra that governs the theory, and some of its representations that  play an important role in the discussion. We also describe the construction of vertex operators that correspond to the holomorphic stress tensor and current algebra generators, and the operators we use for perturbing the theory in later sections.

In sections {\it 3--5}, we discuss the deformation of string theory on $AdS_3\times S^1$ by the operator given by eq. (6.5) in \KutasovXU. This deformation is marginal on the worldsheet but irrelevant in spacetime. The perturbing operator has spacetime scaling dimension $(1,2)$; hence, the corresponding coupling is irrelevant, and breaks Lorentz symmetry. We describe the sigma model of string theory in this background, the semiclassical quantization of this sigma model, and its exact quantization. This leads to a formula for the energies of states in the theory as a function of the above coupling.

In section {\it 6}, we study the problem from a field theoretic point of view. We argue that the theory \aaa\ can be defined such that throughout the RG flow there exists a holomorphic $U(1)$ current $J(x)$, and a holomorphic stress tensor $T(x)$. The spectrum of the resulting theory agrees with that found in string theory. In section {\it 7}, we summarize our main results and discuss possible extensions of this work. Two appendices contain results relevant for the analysis.

{\bf Note added:} after this work was completed, we learned of the related work \ApoloQPQ.

\newsec{Review of string theory on $AdS_3$}

In this section we briefly review some aspects of string theory on $AdS_3$ that will be relevant for our discussion below. We mostly discuss the technically simpler bosonic string, although all the applications we have in mind are in the superstring. The generalization of the formulae below to that case, as well as a more detailed discussion of the bosonic case, appear in \refs{\GiveonNS,\KutasovXU}.

The worldsheet theory on $AdS_3$ is described by the WZW model on the $SL(2,\IR)$ group manifold. This model has an affine $SL(2,\IR)_L\times SL(2,\IR)_R$ symmetry; we will denote the left and right-moving currents  by $J_{\rm{SL}}^a(z)$ and $\bar{J}_{\rm{SL}}^a(\bar{z})$, $a=3,\pm$ , respectively.\foot{In the literature, the  $SL(2,\IR)_L\times SL(2,\IR)_R$ currents are usually denoted by $J^a(z)$, $\bar J^a(\bar z)$. We will use a different notation to distinguish them from the spacetime current $J(x)$ below.}

The level of the worldsheet current algebra, $k$, determines many properties of the model. In the bosonic case, the central charge of the worldsheet CFT is given by $c=3k/(k-2)$. As $k\to\infty$, the central charge goes to three, and the worldsheet sigma model on $AdS_3$ becomes weakly coupled. The radius of curvature of anti de Sitter space is given by $R_{\rm AdS}=\sqrt{k}l_s$, where $l_s=\sqrt{\alpha'}$ is the string length.

The Lagrangian of the WZW model on $AdS_3$ is given by
\eqn\adsact{\LL=2k\left(\partial \phi \bar{\partial}\phi +e^{2\phi}\partial\bar{\gamma}\bar{\partial}\gamma\right),}
where $\phi$ labels the radial direction of $AdS_3$ and $\gamma$ and $\bar{\gamma}$ are complex coordinates on the boundary of $AdS_3$. The background \adsact\ contains both the anti-de-Sitter metric, and a $B$-field, $B_{\gamma\bar\gamma}\sim e^{2\phi}$. The dilaton is constant.

In studying the sigma model on $AdS_3$, \adsact, it is convenient to introduce the auxiliary complex variables $(x,\bar{x})$, \refs{\ZamolodchikovBD,\TeschnerFT}. These variables are especially useful in string theory, where they become the coordinates on the base space of the boundary $CFT_2$.

In terms of these auxiliary variables, the $SL(2,\IR)_L$  currents can be assembled into a single object $J_{\rm{SL}}(x;z)$,
\eqn\curJ{J_{\rm{SL}}(x;z)=2xJ^3_{\rm SL}(z)-J^+_{\rm SL}(z)-x^2J^-_{\rm SL}(z),}
and similarly for the right-moving currents. The $SL(2,\IR)_L$ affine Lie algebra can be written as
\eqn\Jope{J_{\rm{SL}}(x;z)J_{\rm{SL}}(y;w)\sim k\frac{(y-x)^2}{(z-w)^2}+\frac{1}{z-w}\left[(y-x)^2\partial_y-2(y-x)\right]J_{\rm{SL}}(y;w).}
A natural set of primaries of the current algebra corresponds to eigenstates of the Laplacian on $AdS_3$, $\Phi_h(x,\bar x;z,\bar z)$, which take the form
\eqn\eflap{\Phi_h(x;z)=\frac{1}{\pi}\left(\frac{1}{|\gamma-x|^2e^\phi+e^{-\phi}}\right)^{2h}.}
In the quantum theory, these operators give rise to primaries of the worldsheet Virasoro algebra, with scaling dimension $\Delta_h=\bar{\Delta}_h=-h(h-1)/(k-2)$. They are also primaries of $SL(2,\IR)_L\times SL(2,\IR)_R$, with
\eqn\transphih{J(x;z)\Phi_h(y,\bar y;w,\bar w)\sim{1\over z-w}\left[(y-x)^2\partial_y+2h(y-x)\right]\Phi_h(y,\bar y;w,\bar w),}
and similarly for the right-movers.

The relation between $(x,\bar x)$ and position on the boundary can be made explicit by studying the behavior of the primaries  \eflap\ near the boundary, \ie\ as $\phi\to\infty$. One has
\eqn\behphih{\Phi_h={1\over 2h-1}e^{2(h-1)\phi}\delta^2(\gamma-x)+O\left(e^{2(h-2)\phi}\right).}
Thus, operators containing $\Phi_h$ in string theory on $AdS_3$ give rise to local operators in the dual $CFT_2$.

The Lagrangian \adsact\ is difficult to analyze near the boundary of $AdS_3$, $\phi\to\infty$. To study that region, it is convenient to introduce new worldsheet fields $(\beta,\bar\beta)$ and rewrite the Lagrangian in the Wakimoto form \refs{\WakimotoGF \BernardIY-\BershadskyIN},
\eqn\adsWact{\LL=2k(\partial \phi \bar{\partial}\phi +\beta\bar{\partial}\gamma+\bar{\beta}\partial \bar{\gamma}-e^{-2\phi}\beta\bar{\beta}).}
Integrating out $\beta$ and  $\bar{\beta}$ gives back the Lagrangian in eq. \adsact.

Rescaling the fields and treating carefully the measure of the path integral, the Lagrangian \adsWact\ takes the form
\eqn\qWlag{\LL=\partial \phi \bar{\partial}\phi +\beta\bar{\partial}\gamma+\bar{\beta}\partial \bar{\gamma}-\exp\left(-\sqrt{\frac{2}{k-2}}\phi\right)\beta\bar{\beta}-\sqrt{\frac{2}{k-2}}\hat{R}\phi,}
where $\hat{R}$ is the worldsheet curvature; the last term in \qWlag\ indicates that in the Wakimoto formulation the dilaton depends linearly on $\phi$. The string coupling goes to zero as $\phi\to\infty$ like $g_s\sim\exp\left(-\sqrt{\frac{1}{2(k-2)}}\phi\right)$. The interaction term in \qWlag\ goes to zero as well. Thus, near the boundary, the Wakimoto Lagrangian becomes free. This is useful for calculations, and we will use the Wakimoto description below.

String theory on $AdS_3$ is dual to a $CFT_2$; hence, it contains an operator $T(x)$ that gives rise to the holomorphic stress-tensor of the dual CFT. This operator was constructed in \KutasovXU; it has the form
\eqn\stT{T(x)=\frac{1}{2k}\int d^2z(\partial_xJ_{\rm{SL}}\partial_x\Phi_1+2\partial_x^2J_{\rm{SL}}\Phi_1)\bar{J}_{\rm{SL}}(\bar{x};\bar{z}).}
This operator can be thought of as the vertex operator of a particular, almost pure gauge, mode of the graviton-dilaton system on $AdS_3$. It is holomorphic on the boundary (\ie\ $\partial_{\bar{x}} T=0$), has spacetime scaling dimension $(2,0)$, and satisfies the standard OPE algebra of the stress tensor in the spacetime CFT. Flipping all the chiralities  in \stT, $(x,z,J_{\rm{SL}})\leftrightarrow (\bar{x},\bar{z},\bar{J}_{\rm{SL}})$,  gives the anti-holomorphic component of the stress tensor, $\bar{T}(\bar{x})$.

As discussed in \refs{\KutasovXU,\GiveonNIE}, the above worldsheet construction leads to two natural operators with the quantum numbers of $T\bar T$. One is the product of the vertex operator for $T$, \stT, and the analogous one for $\bar T$. This is a double trace operator -- the product of two integrals over the worldsheet. A second one is the vertex operator
\eqn\stD{D(x,\bar x)=\int d^2z(\partial_xJ_{\rm{SL}}\partial_x+2\partial_x^2J_{\rm{SL}})
(\partial_{\bar x}\bar J_{\rm{SL}}\partial_{\bar x}+2\partial_{\bar x}^2\bar J_{\rm{SL}})
\Phi_1(\bar{x};\bar{z}).}
This operator transforms under the left and right-moving boundary Virasoro symmetries generated by $T(x)$ and $\bar T(\bar x)$ as a quasi-primary operator of dimension $(2,2)$; its OPE's with $T$ and $\bar T$ are the same as that of $T\bar T$, but the two operators are distinct. $D(x,\bar x)$ is a single trace operator -- it is a massive mode of the dilaton gravity sector of string theory on $AdS_3$. As shown in \GiveonNIE, adding this operator to the Lagrangian of the boundary theory is the same as adding the operator $J^-\bar J^-$ to the Lagrangian of the worldsheet theory. This perturbed theory was studied in detail in \refs{\GiveonNIE,\GiveonMYJ,\AsratTZD,\GiribetIMM,\ChakrabortyKPR}. On the other hand, adding the operator $T\bar T$ to the Lagrangian of the boundary theory corresponds to a double trace deformation; it was studied in \refs{\McGoughLOL,\ShyamZNQ,\KrausXRN\CardySDV\CottrellSKZ\AharonyVUX\DubovskyDLK\BzowskiPCY\BonelliKIK-\BaggioGCT}.

In a string background of the form $AdS_3\times\NN$, holomorphic currents in the worldsheet theory on $\NN$ give rise to holomorphic currents in the dual CFT \refs{\GiveonNS,\KutasovXU}. Given a holomorphic dimension $(1,0)$ worldsheet $U(1)$ current $K(z)$, one can construct a holomorphic dimension $(1,0)$ spacetime $U(1)$ current\foot{Not to be confused with the worldsheet current $J_{\rm SL}(x;z)$ \curJ.}
\eqn\gcurJ{J(x)=-\frac{1}{k}\int d^2z K(z) \bar{J}_{\rm{SL}}(\bar{x};\bar{z})\Phi_1(x,\bar{x};z,\bar{z}).}
Multiplying this vertex operator by that of $\bar T(\bar x)$ (the anti-holomorphic analog of \stT) gives rise to a double trace deformation of the boundary CFT, analogous to the $T\bar T$ one discussed above.

Just like in that case, there is a single trace analog of the operator $J\bar T$,
\eqn\stA{A(x,\bar{x})=\int d^2 z K(z)(\partial_{\bar{x}}\bar{J}_{\rm{SL}}\partial_{\bar{x}}\Phi_1+2\partial^2_{\bar{x}}\bar{J}_{\rm{SL}}\Phi_1).}
Using the techniques of \KutasovXU\ one can show that this operator has dimension $(1,2)$ in the boundary theory and it transforms under the affine Lie algebra generated by $J(x)$ \gcurJ\ and under Virasoro like the operator $J\bar T$. As explained in \KutasovXU, this operator is not $J\bar T$ since the latter is a double trace operator and \stA\ is a single trace one. As mentioned in the previous section, the relation between the two is similar to that between the operators $\sum_{i,j=1}^N J_i\bar T_j$ and $\sum_{i=1}^N J_i\bar T_i$ in the symmetric product CFT $\MM^N/S_N$.

In this paper we will study the theory obtained by adding to the Lagrangian of the boundary theory the operator \stA. This is an analog of the deformation \stD\ for the $J\bar T$ case. In the $T\bar T$ case, it was shown in \GiveonNIE\ that adding $D(x,\bar x)$ to the Lagrangian of the boundary theory is equivalent to adding the operator $J^-_{\rm SL}\bar J^-_{\rm SL}$ to the Lagrangian of the worldsheet theory. One can repeat the calculation for the $J\bar T$ case, and find
\eqn\marg{\int d^2x A(x,\bar{x})\simeq \int d^2z K(z)\bar{J}^-_{\rm SL}(\bar{z}).}
Thus, adding the operator $A(x,\bar x)$ to the Lagrangian of the boundary theory is equivalent to adding the operator $K\bar J^-_{\rm SL}$ to the worldsheet Lagrangian. In the next section we turn to an analysis of this deformation.

\newsec{$K\bar J^-_{\rm SL}$ deformation $I$ --  worldsheet sigma model}

In this section we discuss string theory on $AdS_3\times S^1$, deformed by adding to the worldsheet Lagrangian the operator $K(z)\bar J^-_{\rm SL}(\bar z)$ (see \marg). Here $K(z)$ is the holomorphic current associated with the left-moving momentum on $S^1$. We denote the coordinate on $S^1$ by $y$, such that
\eqn\curK{K(z)=i\partial y.}
$\bar J^-_{\rm SL}$ is the $SL(2,\IR)_R$ current discussed in the previous section.

We are interested in studying the boundary theory on the cylinder with SUSY preserving boundary conditions, which is dual to the BTZ black hole with $M=J=0$. To construct the background of interest, we start by recalling some properties of the sigma model on massless BTZ$\times S^1$ (see e.g. \ParsonsSI).

An element $g\in SL(2,\IR)$  can be parameterized in Poincar\'e  coordinates as
\eqn\gggg{g=\pmatrix{1 & 0 \cr\gamma & 1 }
\pmatrix{e^{\phi} & 0 \cr 0 & e^{-\phi}}\pmatrix{1 & \bar{\gamma}\cr  0 & 1}
=\pmatrix{e^{\phi} & \bar{\gamma}e^{\phi}\cr
\gamma e^{\phi} & e^{-\phi}+\gamma\bar{\gamma}e^{\phi}}.}
\noindent
The WZW action on $SL(2,\IR)\times U(1)$, $S=S[g,y]$, takes the form\foot{See e.g. \GiveonNS.}
\eqn\wzw{S=\frac{k}{2\pi}\int d^2 z \left(\partial\phi\bar{\partial}\phi+e^{2\phi}\partial\bar{\gamma}\bar{\partial}\gamma
+\frac{1}{k}\partial y\bar{\partial}y\right).}
This action has an affine $SL(2,\IR)_L\times SL(2,\IR)_R\times U(1)_L\times U(1)_R$ symmetry, where the level of $SL(2,\IR)_{L,R}$ is $k$. The massless BTZ black hole is obtained by compactifying the spatial coordinate $\gamma_1$ on the boundary, defined as
\eqn\ttxx{\gamma=\gamma_1+\gamma_0~,\qquad \bar\gamma=\gamma_1-\gamma_0~,}
on a circle of radius $R$,
\eqn\xxrr{\gamma_1\simeq \gamma_1+2\pi R~.}

\noindent
We are interested in a deformation of \wzw\ by the marginal operator
\eqn\dWZW{\delta S=-\frac{\epsilon}{\pi}\int d^2 z K\bar{J}^-_{\rm SL}.}
To first order in $\epsilon$, the deformed action is given by
\eqn\wzwe{S(\epsilon)=\frac{k}{2\pi}\int d^2 z \left(\partial\phi\bar{\partial}\phi+e^{2\phi}\partial\bar{\gamma}\bar{\partial}\gamma+2\epsilon e^{2\phi}\partial y\bar{\partial}\gamma+\frac{1}{k}\partial y\bar{\partial}y\right);}
the dilaton remains constant.

In general one expects higher order corrections in $\epsilon$, however, one can check that in this case the background \wzwe\ is an exact solution of the $\beta$-function equations to leading order in $\alpha'$ (\ie\ in the gravity approximation). In the type II superstring, there are no higher order corrections in $\alpha'$, due to the fact that the background preserves $(2,2)$ worldsheet supersymmetry \BarsSR.

Some additional properties of the background \wzwe\ are:
\item{(1)}
Upon dimensional reduction along the $y$ direction,\foot{Or in the heterotic string, if $K(z)$ is taken from the chiral internal space.}
one obtains the null warped $AdS_3$ background discussed in \refs{\IsraelVV\DetournayFZ - \AzeyanagiZD},
\eqn\dswarp{ds^2=k\left(d\phi^2+e^{2\phi}d\gamma d\bar\gamma-\epsilon^2 e^{4\phi}d\gamma^2\right),}
with a gauge field, $A_\gamma= 2\sqrt{k}\epsilon e^{2\phi}$, and a $B$-field, $B_{\gamma\bar\gamma}=ke^{2\phi}/2$.
\item{(2)} The geometry of the background \wzwe\ has no curvature singularity and, in particular,
the scalar curvature is an $\epsilon$-independent constant, ${\cal R}\simeq -1/k$.
\item{(3)} The worldsheet theory \wzwe\ has an affine
$SL(2,\IR)_L\times U(1)_{R,{\rm null}}\times U(1)_L\times U(1)_R$ symmetry;
the $SL(2,\IR)_R$ symmetry of \wzw\ is broken to $U(1)_{R,{\rm null}}$, associated with translations of $\bar\gamma$,
while the other affine symmetries of the original WZW model are preserved along the deformation line \dWZW.

\noindent
Rescaling $\phi,\gamma,\bar\gamma$ by $1/\ell$, $y$ by $1/\ell_s$, and $\epsilon$ by $\ell_s/\ell$, with
\eqn\stlen{\ell= \sqrt{k}\ell_s; \;\;\; \ell_s=\sqrt{\alpha'},}
the action takes the standard form for a closed string with a cylindrical worldsheet, parametrized
by $\tau$ and $\sigma\simeq\sigma+2\pi$:
\eqn\Lag{{\SS}={1\over 2\pi\alpha'}\int d^2z
\left(\partial\phi\bar{\partial}\phi+e^{\frac{2\phi}{\ell}}\partial\bar{\gamma}\bar{\partial}\gamma+2\epsilon e^{\frac{2\phi}{\ell}}\partial y\bar{\partial}\gamma+\partial y\bar{\partial}y\right),}
where
$z={1\over\sqrt 2}(\tau +\sigma)$,  $\bar{z}={1\over\sqrt 2}(\tau-\sigma)$
and $\partial={1\over\sqrt 2}(\partial_{\tau}+\partial_{\sigma})$,
$\bar{\partial}={1\over\sqrt 2}({\partial}_{\tau}-{\partial}_{\sigma})$.

In the next two sections we will study the spectrum of string theory in the background \Lag, first semiclassically (in the next section), and then exactly (in the following one).

\newsec{$K\bar J^-_{\rm SL}$ deformation $II$: semiclassical analysis of the spectrum}

In this section, as a warmup exercise towards an exact evaluation of the spectrum in the next section, we will perform a semiclassical calculation of the spectrum. In particular, we will take the level $k$ to be large, and perform a semiclassical quantization of the Lagrangian \Lag, keeping only contributions of zero modes.

Thus, we consider a short string, that is located at a particular $\phi$ and carries momentum $P_\phi$ in the $\phi$ direction, energy $E$ conjugate to $\gamma_0$, momentum $P$ conjugate to $\gamma_1$ (see \ttxx), and momentum and winding $n_y$ and $m_y$ in the $y$ direction.

At large $k$ and $\phi\to\infty$, we expect the semiclassical result to match the contribution of the zero modes to the exact dispersion relation, which can be obtained by studying the scaling dimensions of vertex operators in the  worldsheet CFT.

The conjugate momentum densities for the fields $(\phi, \gamma, \bar{\gamma},y)$ are given by
\eqn\momphi{\eqalign{\Pi_{\phi}=&T\dot{\phi},\cr
\Pi_{\gamma}=&{1\over 2}Te^{\frac{2\phi}{\ell}}(\dot{\bar{\gamma}}+\bar{\gamma}')+T\epsilon e^{\frac{2\phi}{\ell}}(\dot{y}+y'),\cr
\Pi_{\bar{\gamma}}=&{1\over 2}Te^{\frac{2\phi}{\ell}}(\dot{\gamma}-\gamma'),\cr
\Pi_{y}=&T\epsilon e^{\frac{2\phi}{\ell}}(\dot{\gamma}-\gamma')+T\dot{y},}}
where  dot and prime denote derivatives with respect to $\tau$ and $\sigma$, respectively,
and $T={1\over 2\pi\alpha'}$ is the string tension.

The worldsheet Hamiltonian,
\eqn\HamilTot{H=\frac{1}{2\pi}\int_0^{2\pi} d\sigma \cal{H},}
is obtained from the Hamiltonian density
\eqn\HamilD{\eqalign{{\cal{H}}= & \Pi_\phi\dot{\phi}+\Pi_\gamma\dot{\gamma}+\Pi_{\bar{\gamma}}\dot{\bar{\gamma}}+\Pi_y\dot{y}-{\cal{L}}\cr
=&\Pi_{\gamma}\gamma'-(\bar{\gamma}'+2\epsilon y')\Pi_{\bar{\gamma}}+\frac{2e^{-\frac{2\phi}{\ell}}}{T}\Pi_\gamma\Pi_{\bar{\gamma}}
+\frac{1}{2T}(\Pi_y-2\epsilon\Pi_{\bar{\gamma}})^2+\frac{Ty'^2}{2}+\frac{\Pi_{\phi}^2}{2T}+{1\over 2}T\phi'^2.}}
The worldsheet momentum is given by
\eqn\momgen{p=\frac{1}{2\pi}\int_0^{2\pi}d\sigma {\cal{P}},}
in terms of the momentum density
\eqn\momdens{{\cal{P}}=\Pi_\phi \phi'+\Pi_{\gamma}\gamma'+\Pi_{\bar{\gamma}}\bar{\gamma}'+\Pi_y y'.}
The spacetime quantum numbers are given by
\eqn\PiQumNo{\eqalign{\int_0^{2\pi} d\sigma \Pi_\gamma &=-\frac{E_L}{R}, \cr
\int_0^{2\pi} d\sigma \Pi_{\bar{\gamma}} &=\frac{E_R}{R},\cr
\int_0^{2\pi} d\sigma \Pi_{y} &=\frac{n_y}{R_y},\cr
\int_0^{2\pi} d\sigma \Pi_{\phi} &=P_\phi,}}
and~\foot{Here and below, w.l.g. we take $w$ to be positive.}
\eqn\winQumNo{\eqalign{\int_0^{2\pi} d\sigma \gamma' &=\int_0^{2\pi} d\sigma \bar{\gamma}'=2\pi wR, \cr
\int_0^{2\pi} d\sigma y' &=2\pi m_yR_y,\cr
\int_0^{2\pi} d\sigma \phi' &= 0,}}
where
\eqn\EL{\eqalign{E_L=&\frac{R}{2}\left(E+P\right),\cr
E_R=&\frac{R}{2}\left(E-P\right).}}
Apart from $P_\phi$, all the charges in \PiQumNo\ -- \EL\ are conserved.

The contribution of the zero modes of the string to the worldsheet energy $H$, \HamilTot, which we denote by $\Delta+\bar{\Delta}$, is thus
\eqn\sumDelta{\eqalign{\Delta+\bar{\Delta}= &-w(E_L+E_R)-\frac{\alpha'}{2} e^{-\frac{2\phi}{\ell}}(E^2-P^2)+\frac{(q_L^2+q_R^2)}{2}  \cr
& +\frac{2\sqrt{2}\epsilon \ell q_LE_R}{R} +\frac{2\epsilon^2 \ell^2 E_R^2}{R^2}+\frac{\alpha'}{2}P_\phi^2~,}}
where $q_L$ and $q_R$ are defined as
\eqn\qL{q_L=\frac{1}{\sqrt{2}}\left(\frac{n_y\ell_s}{R_y}+\frac{m_yR_y}{\ell_s}\right),}
\eqn\qR{q_R=\frac{1}{\sqrt{2}}\left(\frac{n_y\ell_s}{R_y}-\frac{m_yR_y}{\ell_s}\right).}
The contribution of the zero modes of the string to the worldsheet momentum $p$, \momgen, is
\eqn\diffDelta{\eqalign{\bar{\Delta}-\Delta=\frac{1}{2}(q_R^2-q_L^2)+nw,}}
where
\eqn\Pmom{P=\frac{n}{R}~.}
In this section we discussed the semiclassical quantization of the zero modes in the sigma model \Lag. To study string theory in this background we need to go beyond the zero modes, and impose the consistency conditions of string theory on the states, to get the spectrum of the spacetime theory. This is the topic of the next section, where we study the worldsheet and spacetime spectrum of string theory on the deformed massless BTZ$\times S^1$
background \Lag\ exactly.

\newsec{$K\bar J^-_{\rm SL}$ deformation $III$ -- exact analysis of the spectrum}

In this section, we study the spectrum of string theory in the $K\bar J^-_{\rm SL}$ deformed massless BTZ$\times S^1$ background \Lag. We start, in subsection {\it 5.1}, by reviewing the spectrum of the undeformed theory, $M=J=0$ BTZ, which corresponds to a Ramond ground state of the boundary $CFT_2$, following \ParsonsSI. We also review the relation of the boundary $CFT_2$ to a symmetric product CFT, following \refs{\GiveonMI,\GiveonMYJ}.

In subsection {\it 5.2}, we discuss the $J^-_{\rm SL}\bar J^-_{\rm SL}$ deformation on the system of the first subsection.  We rederive some of the results of \refs{\GiveonNIE,\GiveonMYJ,\AsratTZD} from this point of view, and discuss their relation to the symmetric product theory $\MM^N/S_N$ with a $T\bar T$ deformed block $\MM$. In particular, we show that the spectrum found in \refs{\SmirnovLQW,\CavagliaODA} agrees with that obtained from the string theory analysis.

In subsection {\it 5.3}, we use the techniques of the previous two subsections to study the string spectrum in the
$K\bar J^-_{\rm SL}$ deformed massless BTZ$\times S^1$, \Lag. We derive the spectrum of the theory, and in section {\it 6} compare it to the field theoretic analysis of $J\bar T$ deformed $CFT_2$ for the symmetric product theory, $\MM^N/S_N$, with a $J\bar T$ deformed block $\MM$.

\subsec{$M=J=0$ BTZ}

\noindent{\sl $\;\;5.1.1$. The spectrum of the worldsheet theory}

\medskip
To analyze the spectrum of the worldsheet sigma model on the massless BTZ background, it is convenient to use the Wakimoto representation introduced in section {\it 2}, and take the limit $\phi\to\infty$, in which the theory becomes free in these variables. In this limit, the Lagrangian \qWlag\ takes the form\foot{In this section we take $\alpha'=2$,
and choose the periodicity of $\sigma$ such that no factors of $2\pi$ appear.}
\eqn\WLaga{{\cal{L}}_W= \beta\bar{\partial} \gamma + \bar{\beta}\partial\bar{\gamma}+\cal{L}_\phi~,}
where
\eqn\Lphia{{\cal{L}}_\phi=\partial \phi \bar{\partial} \phi-\sqrt{\frac{2}{k}}\hat{R}\phi~.}
Comparing to \qWlag, note that we replaced the factor $k-2$ by $k$, to account for the difference between the bosonic and fermionic string. In \Lphia, $k$ is the total level of the $SL(2,\IR)$ current algebra; it receives a contribution $k+2$ from the bosons and $-2$ from three free fermions (whose contribution to the Lagrangian we did not write).

Following  \ParsonsSI, we bosonize the $\beta-\gamma$ system as
\eqn\Wbosa{\eqalign{\gamma = i\phi_-, &  \ \ \ \ \bar{\gamma} = i\bar{\phi}_-, \cr
\beta=i\partial \phi_+, & \ \ \ \   \bar{\beta}= i\bar{\partial} \bar{\phi}_+,}}
where $\phi_{\pm}$ are given by
\eqn\phipma{\phi_\pm=\frac{1}{\sqrt{2}}(\phi_0\pm \phi_1).}
The fields $\phi_0$ and $\phi_1$ are canonically normalized timelike and spacelike scalar fields, respectively,
\eqn\phiyOPEa{\phi_\mu(z)\phi_\nu(w) \sim -\eta_{\mu\nu}\ln (z-w); \ \ \  \eta_{\mu\nu}={\rm diag}(-1,1); \ \ \ \ \mu,\nu=0,1.}
In terms of the fields $\phi_\pm$, \phipma, the Lagrangian \WLaga\ is given by
\eqn\Wlagfa{{\cal{L}}=-\partial\phi_+\bar{\partial}\phi_--\partial\phi_-\bar{\partial}\phi_+ +{\cal L}_\phi~.}
Vertex operators in the massless BTZ background, correspond to eigenstates of $J^-_{\rm SL}$ and $\bar J^-_{\rm SL}$ with eigenvalues $E_L$ and $E_R$, respectively. In the Wakimoto representation one has
\eqn\Wcura{J^-_{\rm SL}=\beta=i\partial\phi_+;\;\;\;  \bar{J}^-_{\rm SL} = \bar{\beta}=i\bar\partial\bar\phi_+.}
The vertex operators of interest take the form
\eqn\vvvva{V=e^{\sqrt{2\over k}j(\phi+\bar\phi)}V_{E_{L,R}}^w~,}
where
\eqn\veqwa{\eqalign{V_{E_{L,R}}^w  = & e^{iw\phi_+ + iE_L\phi_-}e^{iw\bar{\phi}_+ + iE_R\bar{\phi}_-}\cr
 = & e^{iw\frac{1}{\sqrt{2}}(\phi_0+\phi_1)+i\frac{(E+P)R}{2}{1\over\sqrt{2}}(\phi_0-\phi_1)} \cr
 & \times  e^{iw\frac{1}{\sqrt{2}}(\bar{\phi}_0+\bar{\phi}_1)+i\frac{(E-P)R}{2}{1\over\sqrt{2}}(\bar{\phi}_0
 -\bar{\phi}_1)},}}
with $(E_L,E_R)={R\over 2}(E+P,E-P)$, as in \EL, and
\eqn\vardefa{P  =\frac{n}{R}~.}
$R$ sets the energy scale in the problem. One can think of it as the radius of compactification of the geometric coordinate $\gamma_1$  \xxrr.

The quantum number $j$ in \vvvva\ is related to the momentum in the $\phi$ direction. As discussed earlier in the paper, for the purpose of comparing to the boundary theory we are mainly interested in states carrying real momentum $P_\phi$, which means that $j$ takes the form
\eqn\formjjj{j=-\half+is,\;\;\; s\in \IR,}
with $s$ proportional to $P_\phi$.

The integer $w$ labels different twisted sectors.\foot{One can also think of $w$ as the winding number of the string around the circle on the boundary of BTZ. This is particularly clear in the semiclassical approach of section {\it 4}.} These sectors are constructed in a way analogous to \refs{\MaldacenaHW,\ArgurioTB}, but here the spectral flow/twist is in the $J^-_{\rm SL}$ direction, whereas there it was in the $J^3_{\rm SL}$ direction.

The left and right-moving scaling dimensions of $V$ are
\eqn\DelDelbara{\eqalign{&\Delta  = -wE_L-\frac{j(j+1)}{k}, \cr
&\bar{\Delta} = -wE_R-\frac{j(j+1)}{k},\cr
&\bar{\Delta}-\Delta  =wn.} }
These equations are the exact versions of the semiclassical results \sumDelta, \diffDelta.\foot{The latter also include the contributions from the $S^1$ labeled by $y$.} We will next use them to calculate the spectrum of the spacetime theory.

\bigskip

\noindent{\sl $\;\;5.1.2$. The spectrum of the spacetime theory}

\medskip

Consider the type II superstring on massless  $BTZ\times \cal N$, which corresponds to a Ramond ground state of the dual $CFT_2$. Let
\eqn\verpa{V_{\rm phys}=e^{-\varphi-\bar{\varphi}}V_{BTZ} V_{\cal N}}
be a physical vertex operator of the theory in the $(-1,-1)$ picture,
where $V_{BTZ}$ is a vertex operator in the worldsheet CFT of massless BTZ
and  $V_{\cal N}$ is the vertex operator in the CFT on $\cal N$.

The worldsheet theory on $BTZ\times \cal N$ has ten towers of bosonic and fermionic oscillators (modulu effects that decrease this by an amount of order $1/k$). The physical state conditions in string theory eliminate two towers of oscillators. For our purpose it is sufficient\foot{These are of course not the most general physical vertex operators, since the physical state condition requires the vertex operators to be $N=1$ superconformal primaries in the full theory, whereas we are considering operators that are separately primary in $BTZ$ and in $\NN$, and moreover ones where all the transverse oscillations of the strings are in $\NN$.}  to consider vertex operators whose BTZ component has the form \vvvva, and whose component in $\NN$ has left (right) moving scaling dimension $N$ $(\bar N)$. The left and right scaling dimensions $\Delta$ and $\bar\Delta$ of $V_{BTZ}$ for such states are given in \DelDelbara. The on-shell condition reads:
\eqn\onshella{\eqalign{\Delta+N-\frac{1}{2} &=0~,\cr
\bar{\Delta}+\bar{N}-\frac{1}{2} &=0~.}}
Plugging \DelDelbara\ into \onshella\ leads to the dispersion relations
\eqn\hhha{\eqalign{E_L=\frac{1}{w}\left[-\frac{j(j+1)}{k}+N-\frac{1}{2}\right],\cr
E_R=\frac{1}{w}\left[-\frac{j(j+1)}{k}+\bar{N}-\frac{1}{2}\right];}}
The states that satisfy \hhha\ can be thought of as describing a string that winds $w$ times around the spatial circle in the BTZ geometry, and is moving with a certain momentum (proportional to $s$, \formjjj) in the radial direction, in a particular state of transverse oscillation. Equation \hhha\ gives the energy and momentum of such a state.

It was recognized a long time ago \GiveonMI, that the spectrum \hhha\ is the same as that of a symmetric product CFT, $\MM^N/S_N$, where the block $\MM$ is the $c=6k$ CFT associated with one string in the $BTZ$ background, and the winding $w$ labels the twisted sectors.  The string state \verpa\ corresponds to an operator with dimension $h_w$ in the CFT $\MM^w/Z_w$ which, when acting on the Neveu-Schwartz vacuum of this CFT, creates a Ramond sector state on the cylinder, with energy
\eqn\eellrr{E_L=h_w-{kw\over4};\;\;\; E_R=\bar h_w-{kw\over4}.}
Plugging \eellrr\ into \hhha, we get an equation that can be written as
\eqn\hwhonea{h_w={h_1\over w}+{k\over 4}\left(w-{1\over w}\right),}
which gives the dimensions of the operators in the $Z_w$ twisted sector, $h_w$, in terms of those in the sector with $w=1$, $h_1$. A similar equation holds for the right-movers. Equation \hwhonea\ is a well known expression for the dimension of operators in the $Z_w$ twisted sector of a symmetric product $CFT_2$, $\MM^N/S_N$, where the block $\cal M$ has central charge $c_{\cal M}=6k$ \refs{\KlemmDF,\FuchsVU}.

The qualitative reason to expect a relation between the string spectrum and the symmetric orbifold is the following. As mentioned above, the states \verpa\ can be thought of as describing strings moving in the radial direction in a particular state of excitation. These strings are free (at large $N$, or small string coupling), and the fact that they can be described by a symmetric product is very similar to that utilized in matrix string theory \refs{\MotlTH,\DijkgraafVV}.

The above picture is heuristic. One can view our discussion of deformations of $AdS_3$ below as providing further evidence for it, since we will be able to match field theoretic properties of the symmetric product with a deformed block $\MM$ to the string analysis in the appropriate deformed backgrounds. We discuss these backgrounds in the next two subsections.

\subsec{$J^-_{\rm SL}\bar J^-_{\rm SL}$ deformed massless BTZ}

According to the discussion of the last subsection, fundamental string excitations of the massless BTZ background are described exactly by the symmetric product CFT $\MM^N/S_N$, where $N$ is the number of strings, and $\MM$ the CFT describing one string in this background. One of the interesting aspects of this picture is that, if it is correct, then the deformation by $J^-_{\rm SL}\bar J^-_{\rm SL}$ corresponds to deforming the symmetric product to $\MM_t^N/S_N$, where $\MM_t$ is the CFT $\MM$ deformed by the $tT\bar T$ deformation of \refs{\SmirnovLQW,\CavagliaODA}. Thus, one expects the string theory analysis of the spectrum to yield in this case the same results as that of the spectrum of $T\bar T$ deformed CFT in the above papers.

The fact that this is the case was shown in \GiveonMYJ, however, there it was done indirectly, by working in the Neveu-Schwartz vacuum of the boundary theory and arguing that some results should carry over to the Ramond vacuum. In this subsection we will use the techniques reviewed in the previous one to do the calculation directly in the Ramond sector.

\medskip

\noindent{\sl $\;\;5.2.1$. The spectrum of the worldsheet theory}

\medskip

As discussed in \GiveonNIE, in the Wakimoto representation, the $J^-_{\rm SL}\bar J^-_{\rm SL}$ deformation of the sigma model on $AdS_3$ takes the form (at large $\phi$)
\eqn\llbba{\cal L=\beta\bar\partial\gamma+\bar\beta\partial\bar\gamma-\lambda\beta\bar\beta+\cal L_\phi=
-\partial\phi_+\bar\partial\phi_--\partial\phi_-\bar\partial\phi_++\lambda\partial\phi_+\bar\partial\phi_+
+\cal L_\phi~,}
where we used equations \WLaga--\Wcura. Following \GiveonNIE, we will refer to the deformed background as $\MM_3$.

As we see in \llbba, the deformation acts on the two dimensional space labeled by $(\phi_+,\phi_-)$ by changing the metric $G_{\mu\nu}$ from $\eta_{\mu\nu}$, \phiyOPEa, to
\eqn\ggll{G=\pmatrix{
\lambda & -1\cr
-1 & 0\cr
},}
where we work in the $(+,-)$ basis rather than $(0,1)$ as in \phiyOPEa.

The spectrum of a theory with a general constant metric, such as \ggll, is a familiar problem in string theory, in the context of toroidal compactifications, where it gives rise to the Narain moduli space. The slight novelty here is that the deformation involves time, but we can still use techniques developed in the Narain context, and we will do that below.

After the deformation, the scaling dimensions of the vertex operators \vvvva, \veqwa, which become operators in the sigma model on $\MM_3$, are given by
\eqn\deltaaa{\eqalign{\Delta &=\frac{P_LP_L^t}{2}-\frac{j(j+1)}{k},\cr
\bar{\Delta} &= \frac{P_RP_R^t}{2}-\frac{j(j+1)}{k},}}
where~\foot{See e.g. the review \GiveonFU, around (2.4.12) (with $L\leftrightarrow R$).
The antisymmetric background $B$ is zero in the present example;
we keep it for the case considered in the next subsection.}
\eqn\pLpRa{\eqalign{&P_L =\left(n^t+m^t(B-G)\right)e^\ast,\cr
 &P_R =\left(n^t+ m^t(B+G)\right)e^\ast,\cr
& e^\ast(e^\ast)^t =\frac{1}{2}G^{-1},}}
with~\foot{The light-cone momentum is $(n_+,n_-)={1\over\sqrt 2}(n_0+n_1,n_0-n_1)$,
and similarly for the light-cone winding $m$.}
\eqn\nma{\eqalign{n^t& =(n_+,n_-) =\frac{1}{\sqrt 2}\left(2w, \ ER\right),\cr
m^t & =(m_+,m_-) =\frac{1}{\sqrt 2}\left(P R, \ 0\right).}}
Substituting \pLpRa\ and \nma\ in \deltaaa, we get that operators of the form \vvvva\ in $\MM_3$
have left and right scaling dimensions
\eqn\ddbara{\eqalign{&\Delta  = -wE_L-\frac{\lambda R^2}{8}\left(E^2-P^2\right)-\frac{j(j+1)}{k}, \cr
&\bar{\Delta} = -wE_R-\frac{\lambda R^2}{8}\left(E^2-P^2\right)-\frac{j(j+1)}{k},\cr
&\bar{\Delta}-\Delta  = wn,} }
with $E_{L,R}$ given in terms of the energy and momentum $E,P$ and the radius $R$ in \EL. This equation generalizes \DelDelbara\ to $\lambda\not=0$.

\medskip

\noindent{\sl $\;\;5.2.2$. The spectrum of the spacetime theory}

\medskip

We can use the results of {\sl 5.2.1} to calculate the spectrum of the spacetime theory, as we did in the undeformed, $\lambda=0$, case above. Using the mass-shell condition \onshella, we find the dispersion relation
\eqn\hbarheea{\eqalign{h_w-\frac{kw}{4}=& E_L+\frac{\lambda R^2}{8w}\left(E^2-P^2\right),\cr
\bar{h}_w-\frac{kw}{4}=& E_R+\frac{\lambda R^2}{8w}\left(E^2-P^2\right),\cr
h_w-\bar{h}_w=& n.}}
where $h_w$, $\bar h_w$ are properties of the undeformed theory (\eg, they can be obtained by setting $\lambda=0$ in \hbarheea, and using the dispersion relations of the undeformed theory, \hhha).

It is interesting to compare the spectrum \hbarheea\ to the field theory analysis of \refs{\SmirnovLQW,\CavagliaODA}. It is easy to see that the two agree, if we take the boundary $CFT_2$ to be the symmetric product $\MM^N/S_N$, interpret the deformation \llbba\ to be the $T\bar T$ deformation in $\MM$, and take the coupling $\lambda$ in the string theory problem to be related to the $T\bar T$ coupling, $t$, via
\eqn\tlrr{t= {\pi\over2}\lambda R^2.}
For $w=1$, the spectrum \hbarheea\ is just that of \refs{\SmirnovLQW,\CavagliaODA} for the deformed block $\MM$, $\MM_t$. For $w>1$, it is that of the $Z_w$ twisted sector of the symmetric product $\MM_t^N/S_N$ \GiveonMYJ. 

\subsec{$K\bar J^-_{\rm SL}$ deformed massless BTZ$\times S^1$}

We are now ready to discuss the problem of interest in this paper, the single trace deformation introduced in section {\it 3}. We repeat the steps of the previous two subsections to compute the spectrum, and then, in section {\it 6}, compare to the field theory analysis.

\medskip

\noindent{\sl $\;\;5.3.1$. The spectrum of the  worldsheet theory}

\medskip

The deformed $SL(2,\IR)\times U(1)$ sigma model Lagrangian \wzwe\ in Wakimoto variables is given by\foot{In this subsection we set the radius of the $y$ circle to the self-dual radius, $R_y/\ell_s=1$, for simplicity. The coupling $\lambda$ is proportional to $\epsilon$ in \wzwe.}
\eqn\WLag{{\cal{L}}_W= \beta\bar{\partial} \gamma + \bar{\beta}\partial\bar{\gamma}+\partial y\bar{\partial}y
-2\lambda K\bar{J}^-+{\cal{L}}_\phi~,}
where ${\cal L}_\phi$ is given in \Lphia, and the right-moving $SL(2,\IR)$ current $\bar{J}^-_{\rm SL}$ is given in \Wcura.
The left-moving $U(1)$ current $K$ is given by  \curK, where $y$ is a canonically normalized scalar,
\eqn\yyya{y(z)y(w)\sim -\ln|z-w|^2~.}
In terms of the fields $\phi_\pm$, defined in \Wbosa--\phiyOPEa, and $y$, the worldsheet Lagrangian at large $\phi$ takes the form
\eqn\Wlagf{{\cal{L}}=-\partial\phi_+\bar{\partial}\phi_--\partial\phi_-\bar{\partial}\phi_+ +\partial y\bar{\partial}y+2\lambda \partial y\bar{\partial}\phi_+ +{\cal L}_\phi~.}
This background involves a non-trivial metric and $B$-field background in the three dimensional spacetime labeled by $(\phi_+,\phi_-,y)$,
\eqn\GB{G=\pmatrix{
0 & -1& \lambda \cr
-1 & 0 & 0\cr
\lambda & 0 & 1
}, \ \ \ \ B=\pmatrix{
0 & 0 & -\lambda \cr
0 & 0 & 0 \cr
\lambda & 0 & 0
}.}
The analogs of \vvvva, \veqwa\ for this case are vertex operators in the deformed massless BTZ$\times S^1$ background, which take the form
\eqn\vvvv{V=e^{\sqrt{2\over k}j(\phi+\bar\phi)}V_{E_{L,R};q_{L,R}}^w~,}
where
\eqn\veqw{\eqalign{V_{E_{L,R};q_{L,R}}^w  = & e^{iw\phi_+ + iE_L\phi_- +iq_Ly}e^{iw\bar{\phi}_+ + iE_R\bar{\phi}_- +iq_R\bar{y}}\cr
 = & e^{iw\frac{1}{\sqrt{2}}(\phi_0+\phi_1)+i\frac{(E+P)R}{2}{1\over\sqrt{2}}(\phi_0-\phi_1)+i\frac{1}{\sqrt 2}(n_y-m_y)y} \cr
 & \times  e^{iw\frac{1}{\sqrt{2}}(\bar{\phi}_0+\bar{\phi}_1)+i\frac{(E-P)R}{2}{1\over\sqrt{2}}(\bar{\phi}_0
 -\bar{\phi}_1)+i\frac{1}{\sqrt 2}(n_y+m_y)\bar{y}},}}
with $(E_L,E_R)={R\over 2}(E+P,E-P)$, as before, and
\eqn\vardef{
q_L =\frac{1}{\sqrt{2}}(n_y-m_y), \ \   q_R = \frac{1}{\sqrt{2}}(n_y+m_y), \ \ P  =\frac{n}{R},  }
where  $n,n_y,m_y\in Z$.

Using standard techniques,
the left and right-moving scaling dimensions of $V$ are given by eqs. \deltaaa, \pLpRa,
with~\foot{Recall that the light-cone momentum is $(n_+,n_-)={1\over\sqrt 2}(n_0+n_1,n_0-n_1)$,
and similarly for the light-cone winding $m$.}
\eqn\nm{\eqalign{n^t& =(n_+,n_-,n_y) =\frac{1}{\sqrt 2}\left(2w, \ ER, \ \sqrt{2}n_y\right),\cr
m^t & =(m_+,m_-,m_y) =\frac{1}{\sqrt 2}\left(P R, \ 0, \  \sqrt{2}m_y\right).}}
Substituting \pLpRa\ and \nm\ in \deltaaa, we get
\eqn\DelDelbar{\eqalign{&\Delta  = -wE_L+{1\over 2}\left(q_L+\lambda E_R\right)^2-\frac{j(j+1)}{k}, \cr
&\bar{\Delta} = -wE_R+\lambda q_L E_R+{1\over 2}\lambda^2 E_R^2+\frac{q_R^2}{2}-\frac{j(j+1)}{k},\cr
&\bar{\Delta}-\Delta  =\frac{1}{2}(q_R^2-q_L^2)+wn.} }

Comparing \DelDelbar\ to \sumDelta, \diffDelta\ in the limit $\phi\to\infty$, we see that the two are identical, if we identify
\eqn\lelr{\lambda={\sqrt 2\epsilon\ell_s\over R}~,}
and $s$ \formjjj\ given by $s=\ell P_\phi/\sqrt2$, where $\ell$ is given in \stlen.

The spectrum of strings, in the sector with $w=0$,
on warped $AdS_3\times S^3$ of the type \dswarp,
was found (using different methods) in \AzeyanagiZD;
their results are in harmony with \DelDelbar.

\medskip

\noindent{\it $\;\;5.3.2$. The spectrum of the spacetime theory}

\medskip

Consider the type II superstring on massless  $BTZ\times S^1\times \cal N$ and let
\eqn\verp{V_{\rm phys}=e^{-\varphi}e^{-\bar{\varphi}}V_{BTZ\times S^1} V_{\cal N}}
be a physical vertex operator of the theory in the $(-1,-1)$ picture, where $V_{BTZ\times S^1}$ is a vertex operator in the CFT on  massless $BTZ\times S^1$ of the sort studied in \vvvv, \veqw, and  $V_{\cal N}$ is the vertex operator in the CFT on $\cal N$.

The scaling dimensions of the operator $V_{BTZ\times S^1}$ in the $\lambda$ deformed massless $\rm BTZ \times S^1$ are given in \DelDelbar. The mass-shell condition for \verp\ takes again the form \onshella, where $N$, $\bar N$ are the left and right-moving scaling dimensions of $V_{\cal N}$.

For $\lambda=0$, this equation takes the form
\eqn\hhh{\eqalign{h_w-\frac{kw}{4}=\frac{1}{w}\left[-\frac{j(j+1)}{k}+\frac{q_L^2}{2}+N-\frac{1}{2}\right],\cr
\bar{h}_w-\frac{kw}{4}=\frac{1}{w}\left[-\frac{j(j+1)}{k}+\frac{q_R^2}{2}+\bar{N}-\frac{1}{2}\right].}}
For general $\lambda$, one has
\eqn\enrrel{\eqalign{h_w-\frac{kw}{4}=& E_L-\frac{\lambda}{w}q_L E_R-\frac{\lambda^2}{2w}E_R^2,\cr
\bar{h}_w-\frac{kw}{4}=& E_R-\frac{\lambda}{w}q_L E_R-\frac{\lambda^2}{2w}E_R^2,\cr
h_w-\bar{h}_w=& n.}}
Note that on the first two lines of \enrrel\ the l.h.s. is independent of $\lambda$, while on the r.h.s. one in general has $\lambda$ dependence from $E_L$, $E_R$.

To understand the physical content of \enrrel, it is convenient to rewrite it as follows. Consider first the case $w=1$, corresponding to a string with winding one. From the discussion in earlier subsections, in this sector we expect to see the spectrum of a $J\bar T$ deformed $CFT_2$ with central charge $c=6k$ (before the deformation).

In this sector, one can write \enrrel\ as follows:
\eqn\hcqq{\eqalign{h_1-\frac{c}{24}-{1\over 2}q_L^2=& E_L(\lambda)-\frac{1}{2}(q_L(\lambda))^2,\cr
E_L-E_R=& n,}}
where $c=6k$, and
\eqn\qller{q_L(\lambda)=q_L+\lambda E_R(\lambda).}
We see that the spectrum is completely determined by the following requirements:
\item{(1)} The quantity $E_L(\lambda)-\frac{1}{2}(q_L(\lambda))^2$ is independent of $\lambda$.
\item{(2)} The charge $q_L$ flows according to \qller.
\item{(3)} For all $\lambda$, $E_L$ and $E_R$ differ by the second line of \hcqq.
\item{(4)} At $\lambda=0$ one has $q_L(0)=q_L$, $E_L(0)=h_1-{c\over 24},\, E_R(0)=\bar h_1-{c\over 24}$.

\noindent
In the next section we will see that the field theory \aaa\ gives the same spectrum, with $\mu=2\lambda R$.

For $w>1$, the spectrum \enrrel\ is compatible with that of a $Z_w$ twisted sector of the symmetric product $\MM^N/S_N$, where the block $\MM$ is deformed as described above \GiveonMYJ.

\newsec{Field theory analysis}

In this section we discuss the theory \aaa\ from a field theoretic perspective. We start with a few simple observations about the undeformed $CFT_2$, obtained by setting $\mu=0$.

Since the theory has a conserved left-moving $U(1)$ current $J(x)$, we can write its left-moving stress tensor as\foot{We take $J(x)$ to be canonically normalized, \ie\ $\langle J(x)J(0)\rangle=1/x^2$.}
\eqn\lefttt{T(x)=\half J^2+T_{\rm coset}(x).}
The first term is the Sugawara stress-tensor associated with the abelian current $J(x)$; it is sensitive to states charged under $J$, and current algebra descendants of all primaries (including uncharged ones). The second term is the stress tensor of the rest of the theory. In particular, $T_{\rm coset}$ commutes with $J$ (their OPE is regular).

As we discuss below, when studying the deformed theory \aaa, it is important to take into account contact terms between operators in the undeformed theory (as in \KutasovXB). In particular, we will need the contact term between $J(x)$ and $\bar T(\bar y)$.  Dimensional analysis says that there are two possible contributions. One is
\eqn\contone{J(x)\bar T(\bar y)=c_1\partial_{\bar x}\delta^2(x-y).}
Such terms appear, for example, in the theory of a scalar field $\Phi$ whose stress tensor has an improvement term (as in Liouville theory). In that case one has
\eqn\liouv{J(x)=i\partial\Phi;\;\;\;\bar T=-\half\left(\bar\partial\Phi\right)^2-Q\bar\partial^2\Phi,}
and $c_1$ \contone\ is proportional to the slope of the dilaton, $Q$. In this case, the charge associated with $J(x)$ is not conserved (see \eg\ \PolchinskiRQ). We will assume that the charge is conserved, and thus not discuss this case here (\ie\ we will set $c_1=0$ below).

The second contact term we may write is
\eqn\contwo{J(x)\bar T(\bar y)=c_2\bar J(\bar x)\delta^2(x-y),}
where $\bar J$ is a right-moving current. Of course, in order for \contwo\ to make sense, such a current needs to exist in the theory, but this is often the case. For example, this is the case in the examples discussed in appendix A, where in general there are multiple $\bar J$'s. The simplest of those examples is the theory \liouv\ with $Q=0$ (the theory of a single scalar field), where $\bar J=i\bar\partial\Phi$. This is also often the case in examples associated with the AdS/CFT correspondence.

As discussed in \KutasovXB, the value of contact terms such as $c_2$ in \contwo\ is not an invariant property of the CFT. Rather, they correspond to a choice of coordinates on the space of theories. In the specific case of interest here, when currents such as $\bar J$ in \contwo\ are present, the space of theories is parametrized in addition to the coupling $\mu$ \aaa\ by the value of the coupling of the marginal operator $J\bar J$, which labels the position on a conformal manifold. What for one choice of contact terms \contwo\ corresponds purely to turning on $\mu$, for another might correspond to a more complicated trajectory in the space of $\mu$ and the moduli.

We would next like to understand the fate of the holomorphic current $J(x)$ and stress tensor $T(x)$ in the theory \aaa\ to leading non-trivial order in $\mu$. We will set the contact terms $c_1$, $c_2$ in the undeformed theory to zero, for simplicity. As discussed above and in appendix A, this corresponds to a choice of coordinates on theory space.

Using the standard OPE's of currents and stress-tensors,
\eqn\opeJT{\eqalign{J(x)J(y)\sim &\frac{1}{(x-y)^2}~,\cr
T(x)J(y)\sim &{J(y)\over (x-y)^2}+{\partial_y J(y)\over x-y}=\partial_y\left(J(y)\over x-y\right),\cr
T(x)T(y)\sim &{c/2\over (x-y)^4}+{2T(y)\over (x-y)^2}+{\partial_y T\over x-y}~,}}
and the relation
\eqn\deltaf{\partial_{\bar x}{1\over x-y}=\pi\delta^2(x-y),}
we find to order $\mu$
\eqn\dbarJTnew{\eqalign{\bar\partial J=&\pi\mu\partial\bar T,\cr
\bar\partial T=&\pi\mu J\partial\bar T,\cr
\partial\bar T=&\pi\mu\left[{c\over12}\bar\partial^3 J+2\bar\partial(J\bar T)-J\bar\partial\bar T\right].}}
The order $\mu$ contributions to $\bar\partial J$, $\bar\partial T$ are proportional to $\partial\bar T$, which vanishes in the undeformed theory. Thus $\bar\partial J$, $\bar\partial T$ vanish to this order. The third equation becomes, at order $\mu$,
\eqn\tbarmu{\partial\bar T=\pi\mu J\bar\partial\bar T.}
We can think of this as the equation
\eqn\thethe{\partial\bar T=\bar\partial\Theta~,\;\;{\rm with} \;\;\Theta=\pi\mu J\bar T.}
This is the familiar fact that adding to the Lagrangian a term that breaks conformal invariance, leads to a non-zero trace of the stress-tensor, proportional to the product of the $\beta$ function and the perturbing operator.

The leading non-vanishing contribution to $\bar\partial J$, $\bar\partial T$ is of order $\mu^2$. It is obtained by substituting \tbarmu\ into \dbarJTnew. Using \thethe\ we find the modified equations
\eqn\modhol{\eqalign{
\bar\partial\left(J-(\pi\mu)^2J\bar T\right)=&0,\cr
\bar\partial\left(T-(\pi\mu J)^2\bar T\right)=&0.
}}
Thus, we see that we can modify the original current $J$ and stress tensor $T$ to
\eqn\jtmu{\eqalign{J(\mu)=&J-(\pi\mu)^2J\bar T+O(\mu^3),\cr
T(\mu)=&T-(\pi\mu J)^2\bar T+O(\mu^3),
}}
such that the modified currents remain holomorphic to order $\mu^2$. Moreover, we see that
\eqn\defcos{T_{\rm coset}=T(\mu)-\half J(\mu)^2}
is independent of $\mu$ to this order.

We now come to the main point of this section. We believe that one can define the theory \aaa\ in such a way that it satisfies the following properties for all $\mu$:
\item{(1)} The $(xx)$ component of the stress tensor, $T_{xx}=T$, remains holomorphic, $\bar\partial T(x;\mu)=0$.
\item{(2)} The current $J(x;\mu)$ remains holomorphic, $\bar\partial J(x;\mu)=0$.
\item{(3)} $T_{\rm coset}$, defined as in \defcos, is independent of $\mu$.

\noindent
There are a number of motivations for the requirements (1)-(3). The first is that we just showed that they are valid to order $\mu^2$. The second is the calculation in appendix A, that shows that they are satisfied classically for a particular class of examples. The third is that (1) and (2) are valid in the string theory analysis of the previous sections (see appendix B), and we will demonstrate shortly that adding (3) leads to the same spectrum as that obtained there.

Condition (3) has an additional motivation that comes from known results in two dimensional gauge theory (see \eg\ \KutasovXQ). One can heuristically think of the original $CFT$ before the deformation \aaa\ as a product of three sectors: the left-moving current sector with stress-tensor $T_J=\half J^2$, the left-moving coset sector with stress tensor $T_{\rm coset}$, \lefttt, and the right-moving sector with stress-tensor $\bar T$. Of course, this is not a direct product, which amounts to the statement that one can write all operators in the $CFT$ as linear combinations of operators that are products of contributions from the three sectors, but the quantum numbers of the different contributions are correlated.

The perturbation \aaa, thought of correctly, mixes the current sector with the right-moving sector, but does not act on the coset. Hence, it is natural to expect that the stress-tensor of the coset, \defcos, is independent of $\mu$. As mentioned above and in appendix A, when the original theory has a conformal manifold, moving around this manifold in general mixes the current sector and the coset, and in order for condition (3) to be correct one has to fine tune the RG trajectory in such a way that this does not happen.

If the theory \aaa\ satisfies conditions (1)--(3), we can compute its spectrum on the cylinder as follow. The fact that $T_{\rm coset}$ is independent of $\mu$ means that in an eigenstate of energy, momentum and charge, $|n\rangle$,  we have
\eqn\fff{\langle n|T_{\rm coset}|n\rangle={\rm independent\; of}\;\mu,}
where we assumed that $\langle n|n\rangle=1$.
Plugging \defcos\ into \fff, we find a relation between the energy, momentum and charge of the state:
\eqn\ggg{RE^L-\half Q^2={\rm independent\; of}\;\mu.}
Here $Q$ is the charge of the state $|n\rangle$, and $E^L$ is its left-moving energy
\eqn\eelldef{E^L=\half(E+P)={E_L\over R}~.}
The second equality relates $E^L$ to the quantity $E_L$ that appears naturally in the string theory analysis in sections {\it 4,5}. Note also that in \ggg\ the two terms separately depend on $\mu$, but the difference does not.

Equation \ggg\ implies a differential equation,
\eqn\diffeqone{R\partial_\mu E^R=Q\partial_\mu Q,}
which was obtained by differentiating \ggg\ w.r.t. $\mu$ and using the fact that $\partial_\mu E^L=\partial_\mu E^R$, which follows from the fact that $E^L$ and $E^R$ differ by $P$, which is quantized in units of $1/R$, and in particular does not depend on $\mu$.

A second differential equation for $E^R$ is the one obtained in \GuicaLIA,
\eqn\hhhhh{\partial_\mu E^R=-\half Q\partial_R E^R.}
Using the fact that the dimensionless quantities $RE^R$ and $Q$ depend on $\mu$ and $R$ only via the combination $\mu/R$, we find that  the charge $Q$ depends on $\mu$ as follows:
\eqn\jjjjj{Q(\mu)=Q(0)+\half \mu E^R(\mu).}
Equations \ggg\ and \jjjjj\ are precisely what we got from the string theory analysis, \hcqq, \qller, in a way that looks quite different, at least superficially.

The spectrum of the theory studied in this section was previously discussed by Guica \GuicaLIA. Our result for the spectrum is different. The origin of the difference is that in \GuicaLIA\ it was assumed that the charge $Q$ that appears in eq. \hhhhh\ (which is eq. (2.23) in \GuicaLIA) is independent of $\mu$, \ie\ it is equal to what we called in \jjjjj\ $Q(0)$. This leads to an energy formula of the form (2.28) in \GuicaLIA, which in our notation has the form
\eqn\compguica{RE^R-\half\mu Q(0) E^R={\rm independent\; of}\;\mu.}
Our result is obtained by substituting \jjjjj\ into \ggg\ (and replacing $E^L$ by $E^R$, which is OK since the difference between the two is independent of $\mu$), which yields
\eqn\ourresult{RE^R-\half\mu Q(0) E^R-{1\over8}\left(\mu E^R\right)^2={\rm independent\; of}\;\mu.}
The extra term in \ourresult\ has a dramatic effect; in particular, it implies that states with sufficiently high energies have the property that their energies become complex when we turn on the interaction. For example, consider states with $Q(0)=0$. If their initial right-moving energy, $E^R_0$, is in the range
\eqn\toolargeE{E^R_0>{2R\over\mu^2}~,}
their energy according to \ourresult\ is complex. It is natural to take the dimensionless parameter $\mu/R$ (which can be thought of as the size of the coupling $\mu$ at the scale $R$) to be small, since in this regime the Kaluza-Klein scale $1/R$ is much lower than the scale $1/\mu$ at which the coupling \aaa\ become strong. Thus, there is a large range of energies, ${1\over R}\ll E\ll {1\over\mu}$, for which the theory is already two dimensional $(E\gg 1/R)$, but the effective coupling at that energy, $E\mu$, is still small.

In the regime $\mu\ll R$, the maximal energy \toolargeE\ is very large, $E^R_{0,\rm max}=2R/\mu^2\gg 1/\mu$, and states that are above this bound are highly excited in the original theory (\ie\ they satisfy $RE^R_0\gg 1$). Thus, while the total number of stable states in the deformed theory is finite, it is very large in this limit.

Another difference between the results for the spectrum \compguica\ and \ourresult\ is that for states that are initially uncharged, $Q(0)=0$, the former implies that their energies do not depend on $\mu$, while the latter says that they do, and in fact, as discussed above, can even become complex.

\newsec{Discussion}

In this paper we discussed $J\bar T$ deformed $CFT_2$ \aaa\ using a combination of field and string theoretic techniques. We argued that one can define the theory in such a way that the left-moving Virasoro symmetry and $U(1)$ affine Lie algebra generated by $T$ and $J$, respectively, are preserved throughout the RG flows, while the right-moving Virasoro symmetry is broken to translations of $\bar x$.

We computed the spectrum of the resulting theory, and showed that the field and string theoretic calculations agree. An important part in this agreement was played by the observation that excitations of the Ramond vacuum of the CFT dual to string theory on $AdS_3$ can be described as Ramond sector states in a symmetric product theory, $\MM^N/S_N$, and the deformation  \marg\ of that string theory corresponds to a $J\bar T$ deformation of the block $\MM$.

The agreement of the resulting spectra, as well as the analogous agreement for $T\bar T$ deformations discussed in \GiveonMYJ\ and in section {\it 5} of this paper, provide further support to the above relation between Ramond sector states in the symmetric product $CFT_2$, $\MM^N/S_N$, and excitations of the dual massless BTZ string vacuum.

There are many natural avenues for further work on the theories discussed in this paper. One of the interesting features of the spectrum we found is that when we turn on the deformation \aaa, states whose energies are above a certain value become unstable (\ie\ their energies develop an imaginary part). For states in the initial theory that are uncharged under the $U(1)$ current \aaa, the upper bound is given in \toolargeE. For general initial charge, $Q(0)$, it is
\eqn\ebound{RE^R_{0,\rm max}=\half\left({2R\over\mu}-Q(0)\right)^2.}
It would be interesting to understand the origin of this phenomenon in field theory and in string theory.

In string theory, this phenomenon is reminiscent of what happens for the $T\bar T$ deformation with ``the wrong sign'' of the deformation parameter. For this sign, the authors of \SmirnovLQW\ say that the theory does not have a vacuum, while those of \CavagliaODA\ show that if the undeformed $CFT_2$ is the free field theory of $n$ scalar fields, the deformed theory is described by the Nambu-Goto action with negative tension. In the string theory description of \refs{\GiveonNIE,\GiveonMYJ,\AsratTZD,\ChakrabortyKPR} one finds in that case a background with a  singularity separating the UV and IR regions of the geometry.

For $T\bar T$ deformations that correspond to double trace operators, it was argued in \refs{\McGoughLOL,\KrausXRN,\CottrellSKZ}, that the deformation corresponds to placing a physical UV cutoff in $AdS_3$. The energetics of $T\bar T$ deformed CFT discussed in \refs{\SmirnovLQW,\CavagliaODA} was interpreted in terms of properties of black holes in the resulting geometry. The existence of a maximal energy corresponds in $AdS$ to the fact that there is a maximal size black hole that fits in the cutoff $AdS$ spacetime.

In the background discussed in this paper, \wzwe, the geometry is non-singular, but the spectrum has similar features to those of wrong sign $T\bar T$. We expect that the interpretation in terms of black holes is similar -- such black holes should have a maximal size, and the upper bound on the energy \ebound\ should be related to that maximal size. It would be interesting to understand these black holes better.

It would also be interesting to study the entanglement entropy (EE) of the theories discussed in this paper, following the recent discussion of the $T\bar T$ deformation case \ChakrabortyKPR. Both in $AdS_3$ and in $\MM_3$, the EE exhibits dependence on the UV cutoff. This is related to the infinite number of states of these theories. It is possible that the finite number of states in the geometry \wzwe\ leads to a finite answer for the EE; it would be interesting to see if that is indeed the case.

Our background, whose spacetime dual has a left-moving conformal and affine $U(1)$ symmetry,
is a particularly tractable example of warped $AdS_3$ backgrounds, that appear
\eg\ in the context of the so called Kerr/CFT correspondence and its extensions,
$3d$ Schrodinger spacetimes and  dipole backgrounds (see e.g. \refs{\CompereJK,\ElShowkCM} and references therein for reviews).
The properties of string theory on the $K\bar J^-_{\rm SL}$ deformed $AdS_3\times S^1$, studied in this paper,
may shed light on some of the physics of the other cases as well. Moreover, it may be of relevance to other backgrounds
in quantum gravity with a finite number of states, such as de Sitter spacetime,
as well as to more general time dependent and/or singular backgrounds, such as the ones discussed in \refs{\ElitzurRT\CrapsII-\DetournayRH}.

The string theory approach to $T\bar T$ and $J\bar T$ deformations suggests that there is a more general set of theories that may enjoy many of the good properties of the two special theories above. In particular, when the $CFT_2$ dual to $AdS_3$ has conserved currents such as $J(x)$, $\bar J(\bar x)$, as well as the conserved, traceless  stress tensor, one can turn on general linear combinations of operators of the form $J\bar J$, $J\bar T$, $T\bar J$, $T\bar T$, etc, and study the theory as a function of the couplings of all these operators. 

From the worldsheet point of view, this corresponds to studying the Narain moduli space of truly marginal perturbations that involve left-moving worldsheet currents such as $K(z)$, $J^-_{\rm SL}(z)$, and right-moving currents $\bar K(\bar z)$, $\bar J^-_{\rm SL}(\bar z)$. The spectrum for a general theory in this class appears in section {\it 5}. For any background $G,B$ that corresponds to a specific combination of $J\bar J$, $J\bar T$, $T\bar J$ and $T\bar T$ in the dual deformed $CFT_2$, one can use eqs. \deltaaa, \pLpRa, and the holographic dictionary, to find the spectrum of the spacetime theory. Thus, the string theory analysis in this paper gives a prediction for the spectrum of the corresponding deformed field theory. It would be interesting to verify it from the field theory point of view.

\bigskip\bigskip
\noindent{\bf Acknowledgements:}
We thank O. Aharony, M. Guica, N. Itzhaki and B. Kol for discussions, and M. Guica for comments on the manuscript. The work of AG is supported in part by the I-CORE Program of the Planning and Budgeting Committee and the Israel Science Foundation (Center No. 1937/12), and by a center of excellence supported by the Israel Science Foundation (grant number 1989/14). The work of DK is supported in part by DOE grant DE-SC0009924. DK thanks the Hebrew University, Tel Aviv University and  the Weizmann Institute for hospitality during part of this work.

\appendix{A}{The $J\bar T$ deformation for scalar fields}

In this appendix we will consider the case where the original $CFT_2$, before the deformation \aaa, is a single free field $\Phi(x,\bar x)$, with the standard Lagrangian
\eqn\llphi{\LL=\frac{1}{4\pi}\partial\Phi\bar\partial\Phi.}
This theory has conserved left and right-moving affine Lie algebras generated by
\eqn\genkm{J(x)=i\partial\Phi;\qquad \bar J(\bar x)=i\bar\partial\Phi.}
The currents $J$ and $\bar J$ are canonically normalized, \ie\ $\langle J(x)J(y)\rangle=1/(x-y)^2$, and similarly for $\bar J$.
The left-moving Virasoro generator can be written in the Sugawara form
\eqn\leftvir{T(x)=\half J^2=-\half \left(\partial\Phi\right)^2,}
and similarly for $\bar T$.

Adding the perturbation \aaa\ to first order in $\mu$ corresponds to adding to the Lagrangian \llphi\ the term
\eqn\newterm{\LL_{\rm int}=\frac{\lambda}{4\pi}\partial\Phi\left(\bar\partial\Phi\right)^2;\;\;\; \lambda=-2i\pi\mu.}
In the presence of the perturbation \newterm, we expect the form of the stress-tensor and current to change, and we have to add higher order terms to the Lagrangian to take this change into account. For the case of the $T\bar T$ deformation, this was done in \CavagliaODA, and for $J\bar T$ it was done in \GuicaLIA\ with slightly different requirements than we will impose. We next mimic their calculations for our case.

To do that, we proceed as follows. The symmetries of the problem suggest that the Lagrangian of the deformed theory takes the form
\eqn\bact{\LL=\frac{1}{4\pi}\partial \Phi\bar{\partial}\Phi\FF(\lambda \bar{\partial}\Phi),}
where $\lambda$ is given in \newterm, and $\FF(z)$ is a function of one variable that satisfies $\FF(0)=\FF'(0)=1$. Our task is to determine this function using two constraints. One is the demand that changing the coupling $\mu$ in \aaa\ corresponds to an insertion of the operator $J\bar T$,
\eqn\flow{\partial_{\mu}\LL=J(x;\lambda)\bar{T}(x;\lambda).}
The second is that the $(xx)$ component of the stress tensor, $T_{xx}=T$, can be written  in the Sugawara form
\eqn\sugform{T=\half J^2.}
Here $J$ is a holomorphic current, whose form is also to be determined.

To solve these constraints, we compute the stress tensor using the Noether procedure,
\eqn\Nst{T_{\mu\nu}=-2\pi\left\{\frac{\delta \LL}{\delta(\partial^\mu \Phi)}\partial_\nu \Phi-g_{\mu\nu}\LL\right\},}
where $g_{xx}=g_{\bar{x}\bar{x}}=0$, $g_{x\bar{x}}=g_{\bar{x}x}=1$, and $\LL=\frac{1}{8\pi}g^{\mu\nu}\partial_\mu \Phi\partial_\nu \Phi\FF(\lambda \bar{\partial}\Phi)$.
This gives
\eqn\compst{\eqalign{T(x;\lambda)= T_{xx} &=-\frac{1}{2}(\partial \Phi)^2\left(\FF+\lambda \bar{\partial}\Phi\FF'\right),\cr
\bar{T}(x;\lambda)=T_{\bar{x}\bar{x}}&=-\frac{1}{2}(\bar{\partial}\Phi)^2\FF,\cr
T_{x\bar{x}} &=- \frac{\lambda}{2}\partial \Phi(\bar{\partial}\Phi)^2\FF',\cr
T_{\bar{x}x} &= 0,}}
 where $\FF'=\partial_z\FF(z)$ is the derivative of $\FF$ w.r.t. its argument. The Euler-Lagrange equation,
 \eqn\eom{\partial\bar{\partial}\Phi\left(\FF+\lambda \bar{\partial}\Phi\FF'\right)+\lambda\partial \Phi\bar{\partial}^2\Phi\FF'+\frac{\lambda^2}{2}\partial \Phi\bar{\partial}\Phi\bar{\partial}^2\Phi\FF''=0,}
is equivalent in this case to holomorphy of the stress-tensor $\bar{\partial}T=0$. Note that this is true for any $\FF$.

Equations \flow, \compst\ give in this case an expression for the current $J(x)$,
\eqn\jz{J(x;\lambda)=i\partial \Phi\frac{\FF'}{\FF}~.}
Furthermore, comparing  the first line of \compst\ with \sugform\ and \jz, we get a differential equation for $\FF(z)$:
\eqn\ffdiff{\FF+z\FF'=-\left(\frac{\FF'}{\FF}\right)^2.}
Together with the boundary condition $\FF(0)=\FF'(0)=1$, the solution is uniquely fixed to be
\eqn\fsolc{\FF(z)=\frac{1}{1- z}~.}
Thus, we see that in the classical theory, the requirements:
\item{(1)}  For all points in the space of field theories labeled by $\mu$ we have a holomorphic current $J(x)$ and a holomorphic stress-tensor $T(x)$;
\item{(2)}  Infinitesimal change of $\mu$ corresponds to the perturbation $J\bar T$ at the point $\mu$, see \flow;
\item{(3)} The holomorphic stress-tensor $T$ can be written in the Sugawara form \sugform,

\noindent
can indeed be satisfied, as expected from the discussion in the text of this paper. Of course, in section {\it 6} we assumed that these requirements are satisfied in the quantum theory as well. The arguments of this appendix are not sufficient to establish that, but we expect this to be true, given the additional evidence described in the text.

In the example discussed here, the coset discussed in the text was empty, \ie\ $T_{\rm coset}=0$. We could generate a non-trivial coset theory by starting with $N>1$ free scalar fields $\Phi_i$, and taking the current $J(x)$ to be a particular left-moving translation current in the $N$ dimensional space,
\eqn\ndimcur{J(x)=i\vec a\cdot\partial\vec\Phi,}
for some $\vec a$. We will not describe the analysis of this case in detail here, but will point out an interesting subtlety in this analysis. Before we turn on the irrelevant deformation \aaa,  this model has an $N^2$ dimensional (Narain) conformal manifold labeled by the coefficients in the Lagrangian of the marginal operators $\partial\Phi_i\bar\partial\Phi_j$. Turning on $\mu$ leads to a complex pattern of flows in which both $\mu$ and the moduli are changing with the scale.

In the field theory analysis this may lead one to think that the structure of the coset depends on $\mu$, in contrast with the claims in the text of this paper. However, this is not the case. The precise statement in theories where the original $CFT_2$ before the deformation \aaa\ has a conformal manifold, is that there exist trajectories in the multi dimensional theory space labeled by the moduli and $\mu$ in which the statement that correlation functions of $T_{\rm coset}$ are independent of $\mu$ is correct.

For example, in the case of multiple scalar fields discussed here, if one fermionized the scalar fields to $N$ complex left and right-moving fermions, the statement would be correct. The difference between the bosonic and fermionic descriptions of the original $CFT$ is in the values of contact terms, and the fact that turning on $\mu$ in one description corresponds to a complicated trajectory in the space of marginal couplings and $\mu$ in the other. The relation between contact terms and reparametrization of theory space is familiar from \KutasovXB. We will leave a more detailed discussion of these issues to another paper.

\appendix{B}{Symmetries of $K\bar J^-_{\rm SL}$ deformed $AdS_3\times S^1$}

Before deforming the worldsheet theory on $AdS_3\times S^1$ by $K\bar J^-_{\rm SL}$, string theory in this background has conserved holomorphic currents $T(x)$, $J(x)$, as well as an anti-holomorphic current $\bar T(\bar x)$ \KutasovXU.
A natural question is what part of this symmetry is preserved by the deformation. We will leave a detailed discussion of this question to future work, but there are a few simple things that can be said already at the level of the Wakimoto representation \WLag.

Consider first the stress tensor $T(x)$, or equivalently the Virasoro generators $L_n$ obtained from $T(x)$ by expanding it in modes, $T(x)=\sum_n L_n x^{-n-2}$. The expressions for the $L_n$ in the Wakimoto representation were given in \GiveonNS, see eqs. (2.36), (3.5). These charges commute with the deformation $K\bar J^-_{\rm SL}$, hence, they are not broken by the perturbation. Thus, one concludes that the deformed theory still has a conserved holomorophic stress-tensor $T(x)$.

Of course, just like in the original $AdS$ background, the Wakimoto analysis is only valid at large $\phi$, and in order to convince oneself that there is indeed a holomorphic stress tensor, one needs to do the analog of \KutasovXU\ for the deformed theory. We will leave this to future work.

We next turn to the $U(1)$ current $J(x)$, which is conserved and holomorphic in the undeformed theory. It is natural to ask whether it remain so after the deformation. As discussed above, this spacetime current is associated with the worldsheet current $K(z)$  (see \gcurJ). It is well known that for worldsheet deformations of the form $J_1(z)\bar J_2(\bar z)$, both currents remain conserved and (anti) holomorphic in the deformed theory. Thus, in our case, in the presence of the deformation $K\bar J^-_{\rm SL}$, $K(z)$ remains holomorphic on the worldsheet. Hence, the corresponding charge, $Q=\oint dz K(z)$, remains conserved. The presence of the spacetime Virasoro generators then implies that the other modes of the current $J(x)$, defined via the expansion $J(x)=\sum_n Q_n x^{-n-1}$, are conserved as well. Again, a more precise treatment would involving generalizing the construction of \KutasovXU\ to this case.

The authors of \AzeyanagiZD\ used the definition of Virasoro generators as graviton vertex operators corresponding to Brown-Henneaux diffeomorphisms parametrized by the left-moving energy, $E_L$, which according to \deBoerGYT\ is equivalent to that of \KutasovXU, to show that $T(E_L)$ satisfy the Virasoro algebra, where $T(E_L)$ is the Fourier transform of $T(x)$, \stT. They argued that this means that the left-moving conformal symmetry in spacetime is unaffected by the deformation.

The authors of \BzowskiPCY\ argued that the right-moving $U(1)$ associated with translations of $\bar\gamma$ might be enhanced to a non-local Virasoro algebra, whose status in the quantum theory is unclear.

\listrefs
\end